\documentclass[fleqn,usenatbib]{mnras}

\usepackage{newtxtext,newtxmath}

\usepackage[T1]{fontenc}

\DeclareRobustCommand{\VAN}[3]{#2}
\let\VANthebibliography\thebibliography
\def\thebibliography{\DeclareRobustCommand{\VAN}[3]{##3}\VANthebibliography}

\usepackage{graphicx}	
\usepackage{amsmath}	
\usepackage{pdflscape}
\usepackage{geometry}
\usepackage{multicol}
\usepackage{enumerate}
\usepackage{epstopdf}
\usepackage{fix-cm}

\newcommand\hi{\mbox{H\,{\sc i}}\ }
\newcommand\hix{\mbox{H\,{\sc i}}}
\newcommand\hiit{\mbox{H\,{\textsl{\textsc{i}}\,}}}

\title[\hiit absorption in star-forming galaxies - II. ]{A search for \hi absorption in distant star-forming galaxies with ASKAP-FLASH - II. Direct observations and stacking of 21\,cm line}

\author[S. L. Eden et al.]{Sophie L. Eden,$^{1}$\thanks{E-mail: sophie.eden@hull.ac.uk}
Elaine M. Sadler,$^{2,3,4}$
Kevin A. Pimbblet,$^{1}$
Stephen J. Curran, $^{5}$
Julia Healy,$^{6,7}$
\newauthor{Filippo M. Maccagni,$^{8,9}$
Elizabeth K. Mahony,$^{3}$
Hyein Yoon,$^{10,2}$ 
}
\\
$^{1}$E. A. Milne Centre for Astrophysics, University of Hull, Cottingham Road, Kingston-upon-Hull, HU6 7RX, UK \\
$^{2}$Sydney Institute for Astronomy, School of Physics A28, University of Sydney, Sydney, NSW 2006, Australia \\
$^{3}$ATNF, CSIRO Space and Astronomy, PO Box 76, Epping, NSW 1710, Australia\\
$^{4}$ARC Centre of Excellence for All Sky Astrophysics in 3 Dimensions (ASTRO 3D) \\
$^{5}$School of Chemical and Physical Sciences, Victoria University of Wellington, Wellington, New Zealand \\
$^{6}$Jodrell Bank Centre for Astrophysics, School of Physics and Astronomy, University of Manchester, Oxford Road, Manchester, M13 9PL, UK\\
$^{7}$United Kingdom SKA Regional Centre (UKSRC), University of Manchester, Oxford Road, Manchester, M13 9PL, UK\\
$^{8}$ASTRON – Netherlands Institute for Radio Astronomy, Oude Hoogeveensedijk 4, 7991 PD Dwingeloo, The Netherlands \\
$^{9}$INAF – Osservatorio Astronomico di Cagliari, via della Scienza 5, 09047 Selargius (CA), Italy \\
$^{10}$Korea Astronomy and Space Science Institute, 776 Daedeokdae-ro, Yuseong-gu, Daejeon 34055, Republic of Korea \\
}

\date{Accepted XXX. Received YYY; in original form ZZZ}

\pubyear{2026}

\begin{document}
\label{firstpage}
\pagerange{\pageref{firstpage}--\pageref{lastpage}}
\maketitle

\begin{abstract}
We present results of a search for associated and intervening  \hi 21\,cm absorption in star-forming galaxies and a control sample of radio galaxies at 0.4\,<\,\textit{z}\,<\,1.0, using ASKAP's FLASH \hi 21\,cm absorption survey. We report the detection of a new \hi 21\,cm absorption line in radio galaxy NVSS J214954-004657 at $z_{\mathrm{HI}}$ = 0.863. Additionally, we present results of \hi 21\,cm stacking for associated and intervening star-forming and radio galaxy catalogues. For the first time at these redshifts, we report tentative detections of stacked \hi 21\,cm absorption for associated galaxies when stacked in order of descending radio source brightness, and show that these tentative detections become buried as optical depth increases as progressively weaker sources are co-added into the stack. No stacked signal is detected when co-adding intervening galaxy spectra. We speculate that the ability to detect a stacked \hi 21\,cm detection is dependent on the background source optical depth limits, as well as source covering factors. Covering factor has a greater impact when stacking intervening systems, and can be considered to effectively weaken the observed flux density of a radio source. We find no significant difference in detection rate for associated \hi 21\,cm absorption in star-forming galaxies versus the general radio-source population at the same redshift. Although detection rates are low, from a meta-analysis of \hi 21\,cm surveys we find that 1 detection is expected given the survey parameters.\end{abstract}




\section{Introduction}\label{introduction}

Reservoirs of cool, neutral gas surround many galaxies in discs \citep{Kaufmann2006,Morganti2006,Wang2020}, including radio galaxies \citep{Emonts2007}. A steady supply of \hi into discs around star forming galaxies appears to be required to sustain reservoirs to allow for ongoing galactic star formation \citep{Larson1980,Keres2005,Sancisi2008,Cortese2021}. The amount of star formation possible in a galaxy is linked to the quantity of \hi in the gas reservoir of a galaxy \citep{Brown2017,Dutta2022,Wagh2024}. Although hydrogen in the form of H$_\mathrm{2}$ is the raw fuel for star formation rather than \hi \citep{Wong2002}, there can be difficulties in tracing star formation via H$_\mathrm{2}$ or CO \citep{Bolatto2013}. Because of this, the 21\,cm \hi line is still often used to trace star formation, as quantities of H$_\mathrm{2}$ are linked to quantities of \hi \citep{Stroe2015}.

Although \hi 21\,cm emission has been used successfully in the local Universe to better understand the history of interactions between the Milky Way and the Magellanic clouds \citep{Putman1998}, as well as those of galaxies nearby to our own such as the M81 group \citep{Yun1994}, the \hi 21\,cm emission line is notoriously faint, being limited to detections for \textit{z}\,<\,0.4 \citep{Fernandez2016,Maddox2021}, except under extraordinary circumstances such as with gravitational lensing \citep{Chakraborty2023}. In order to investigate \hi signals at higher redshifts, absorption signals can be searched for. This requires a geometric setup, where a \hi disc or cloud is positioned with a bright radio continuum source in its background \citep{Brown1973,Wolfe1975,Curran2016b}.

Alternatively, multiple spectra containing no significant \hi detection, can be co-added in a process known as stacking. The resulting stacked signal is an average across all spectra in the stack, and provides information about the average properties of all sources in the stack. For \hi 21\,cm emission, this has successfully revealed emission signals at lower column densities or higher redshifts than a direct detection would require, that could be present in many different spectra. On the other hand, \hi 21\,cm absorption stacking work with galaxies not already known to contain detections of \hi 21\,cm absorption has been unsuccessful.

Historically, \hi 21\,cm absorption work has been attempted at low redshifts below z = 0.25, with galaxies that host active galactic nuclei \citep[AGN;][]{Gereb2014, Maccagni2017}. The first \hi 21\,cm absorption stacking work of its kind, \citet{Gereb2014} shows that if stacking is done with spectra of AGN at 0.02\,<\,\textit{z}\,<\,0.23 where a \hi 21\,cm absorption signal has not been detected, then an absorption signal remains undetected in the stacked spectrum down to very low optical depths ($\tau$ < 0.002), while the average peak optical depth of detected sources is an order of magnitude greater ($\tau_{\mathrm{peak}}$ = 0.02).

\citet{Maccagni2017} continued the \hi 21\,cm absorption stacking work of \citet{Gereb2014} with objects in a slightly wider range of redshifts (0.02\,<\,\textit{z}\,<\,0.25), and similarly found that an absorption signal remains undetected down to an optical depth of $\tau$ = 0.0015. The results of \citet{Gereb2014} and \citet{Maccagni2017} seem to indicate that there are two distinct populations of radio galaxies; one where the galaxies contain \hi which can be detected directly as \hi 21\,cm absorption, and one where if absorption cannot be directly detected, then even stacking the spectra does not reveal a \hi 21\,cm absorption signal. They conclude that in the population of non-detections, while the lack of detection towards some galaxies is likely due to orientation effects, or physical conditions of the gas (high spin temperatures and low column densities), it is possible that some galaxies may be completely depleted of \hix.

\citet{Gereb2014} theorise that as compact (compact steep spectrum; CSS, gigahertz peaked sources; GPS) sources demonstrate higher direct detection rates of \hi 21\,cm absorption compared to extended ones, then this effect could partially be explained by the age of a source. CCS and GPS have been accepted as radio galaxies at earlier stages of their evolution, and as possible precursors to Seyfert galaxies \citep{Fanti1995,ODea1998,Collier2018}. These young, compact AGN show higher rates of direct \hi 21\,cm absorption detection than extended ones \citep{Gereb2014,Maccagni2017}.

In their search at 0.7\,<\,\textit{z}\,<\,1.0, \citet{Murthy2022} detect no evidence of \hi 21\,cm absorption towards any of their 29 radio-loud AGN targets. Further, they do not find a stacked \hi 21\,cm absorption signal. 24 of these targets possess extended radio structures, and they compare this against the rate of direct detections in extended sources at \textit{z}\,<\,0.25. They find that the strength of detections towards extended sources at higher redshifts is much weaker than for extended sources at \textit{z}\,<\,0.25.

The evolution of \hi across redshifts appears to be an important factor in the detection of \hi 21\,cm absorption for both compact and extended sources, potentially caused by either or both neutral gas column density and spin temperature of the gas evolving over cosmic time \citep{Curran2008,Curran2012b,Kanekar2014,Aditya2016}. This poses an issue for searches for detections in these higher redshift ranges.

In the preceding paper in this series \citep{Eden2025}, we outlined the creation of a list of UV-selected target galaxies, containing WiggleZ galaxies \citep{Drinkwater2010,Drinkwater2018}, and an additional sample of galaxies selected to match the selection criteria of the WiggleZ galaxies. WiggleZ galaxies are specifically selected to be star-forming, and thus are expected to have reservoirs of \hi for fuel, making them good targets to search for direct and intervening \hi 21\,cm absorption, as well as for absorption stacking experiments with the First Large Absorption Survey in \hi \citep[FLASH;][]{Allison2022}. We outlined the process of producing two catalogues of optical-radio cross-matches from these WiggleZ-type galaxies and FLASH radio sources, where `associated' galaxies have a <5\,arcsec separation from a FLASH source, which will allow us to probe the medium within the galaxy, while `offset'galaxies have a separation between 5\,arcsec and 20\,arcsec, which will allow us to probe the circumgalactic medium (CGM) surrounding them.

We carried out a series of spectral line analyses of these two samples of WiggleZ-type galaxies, which showed that associated galaxies are 2--5 times more likely to host an active galactic nucleus (AGN), while the offset sample was found to be 80 per\,cent star-forming (SF). We showed why these two catalogues of optical-radio cross-matched objects would make good targets for a search for \hi 21\,cm absorption search at 0.4\,<\,\textit{z}\,<\,1.0.

In this work, we now use these associated and offset galaxy samples to search for \hi 21\,cm absorption. The brightness of the continuum source sets the detection limit for \hi 21\,cm absorption, and so we impose a minimum flux density of $\geq{30}$\,mJy for a source to be considered `bright'. This is because for FLASH, the spectral-line noise for a 2-hour observation is $\sim{5.5}$\,mJy\,bm$^{-1}$\,ch$^{-1}$ \citep{Yoon2025}, which for a 5$\upsigma$ detection would imply a minimum peak optical depth of $\uptau\sim{2.5}$ for a 30\,mJy source. As lines of this strength are very unlikely \citep{Yoon2025}, FLASH does not search for lines in sources fainter than 30\,mJy. If associated or offset objects are cross-matched with a bright FLASH source, we look for direct detections in the FLASH spectra at the redshift priors provided by the optical galaxy spectroscopic redshift. Stacking is then done using all bright galaxies without a direct detection (non-detections), and all `non-bright' (<\,30\,mJy) galaxies, and we look for evidence of \hi in these galaxies that could not otherwise have been detected.

Section \ref{objects} briefly outlines the selection process of our extended WiggleZ sample, as was produced in \citet{Eden2025} in this series, in accordance with WiggleZ selection criteria \citep{Drinkwater2010}, as well as why we suspect these objects are well-placed for an investigation of \hi around galaxies at intermediate redshifts. We also describe the qualities of the associated and offset samples, and how these two different samples allow for targeting of \hi 21\,cm absorption in two different ways. In Section\,\ref{flash_sources} we describe the FLASH survey data used in this work, including the procedures used to obtain spectra for both bright and non-bright objects. We also outline the assumptions used in this work for the conversion of flux density into optical depth. In Section\,\ref{direct_detections} we present an associated detection, as well as an analysis of the galaxy it was detected in. In Section\,\ref{stacking}, we outline the process of stacking, and then investigate the results of stacking FLASH radio spectra taken from the positions of the cross-matched FLASH sources for detections. We also carry out a bootstrapping analysis to help explain the meaningfulness of our findings in the context of the star-forming galaxies we are investigating. Section\,\ref{discussion} and \ref{conclusion} contain a discussion of the findings within this paper, and the conclusions we draw from our findings respectively.
Throughout this work, we assume a concordance cosmology model where $\upOmega_{\upLambda}$\,=\,0.7, $\upOmega_{\mathrm{m}}$\,=\,0.3, and H$_{0}$\,=\,70\,km\,s$^{-1}$\,Mpc$^{-1}$.


\section{Sample object selection}\label{objects}

\subsection{Optical components}\label{wigglez_galaxies}

The WiggleZ survey \citep{Drinkwater2010,Drinkwater2018} is a unique spectroscopic redshift survey designed to observe UV-bright galaxies for the purpose of studying baryonic acoustic oscillations, specifically in the redshift range 0.2\,<\,\textit{z}\,<\,1.0, thus covering the redshifts over which FLASH can detect \hi 21\,cm absorption. It covers five of the equatorial Galaxy and Mass Assembly \citep[GAMA;][]{Driver2011} survey fields, and possesses GALEX UV photometry, as well as Sloan Digital Sky Survey \citep[SDSS;][]{Adelman-McCarthy2006} and Red-sequence Cluster Survey 2 \citep[RCS2;][]{Yee2007} optical photometry, allowing colour cuts to be put in place to select suitable high-redshift UV-bright (NUV\,<\,22.8\,mag) target objects. The survey has a spectral resolution of $R\sim{1300}$, which equates to a dispersion of 0.11\,nm\,pixel$^{-1}$ \citep{Drinkwater2010}, and \citet{Drinkwater2018} report that the redshift uncertainty is 48\,km\,s$^{-1}$ or better for all spectra with QOP redshift quality flags of 3\,--\,5 (single line identifications to multiple line identifications). 90 per\,cent of galaxies in the WiggleZ survey were found to be star-forming. Detail on the suitability of the WiggleZ galaxies for our work is outlined in Section\,2 of \citet{Eden2025}.

\begin{table}
\centering
\begin{tabular}{l|r|r|r|r|r|r}
    \hline \hline
    Field & RA$_{\mathrm{min}}$ & RA$_{\mathrm{max}}$ & Dec$_{\mathrm{min}}$ & Dec$_{\mathrm{max}}$ & Area & N\textsubscript{$\upSigma$(field)}\\
     & (deg) & (deg) & (deg) & (deg) & (deg$^{2}$)\\
    \hline    
    W09 & 133.7 & 150.0 & -0.9 & 3.0 & 23.2 & 5813\\
    W15 & 211.0 & 223.5 & -2.7 & 3.7 & 69.5 & 18388\\
    Ext & 316.5
 & 43.5 & -3.1 & 3.1 & 122.3 & 7462\\
    \hline
\end{tabular}
\caption[Position, area, and total objects contained per field (duplicate from \citet{Eden2025})]{Position on sky, areas of field regions, and total objects fulfilling WiggleZ selection criteria contained within them. Note that the extended sample field wraps around 0h. The total coverage of the sky in this work is 215\,deg$^{2}$. Table is from \citet{Eden2025}.}
\label{tab:areas_table}
\end{table}

We use WiggleZ targets with SDSS photometry from the 9h and 15h fields - these will be referred to as the `W09' and `W15' fields hereon. In \citet{Eden2025} we discussed our interest in extending our sample of WiggleZ-type galaxies to increase the number of potential \hi containing objects to \color{black} be probed at 0.4\,<\,\textit{z}\,<\,1.0. In order to do so, \color{black}we used the WiggleZ selection criteria \citep[outlined in Section 3.5 of][]{Drinkwater2018} to construct an additional field of WiggleZ-type objects, hereon referred to as either the `extended field', or `Ext'.

This was constructed using GALEX FUV and NUV data, DESI Legacy Survey DR9 \citep{Dey2019} spectroscopic redshift data, limited to a redshift slice of 0\,<\,\textit{z}\,<\,1.0 to match the range of the WiggleZ data, and Dark Energy Survey \citep[DES;][]{Abbott2021} \textit{gri} band photometry. This resulted in an equatorial field extending from 21h to 3h, wrapping around 0h. The areas of the fields of objects used in this work are stated in Table \ref{tab:areas_table}. WiggleZ objects which use SDSS \textit{ugriz} band photometry are magnitude-limited and have a completeness of 95 per\,cent to 22.0, 22.2, 22.2, 21.3 and 20.5\,mag, respectively \citep{Drinkwater2010}. Similarly, objects in the Ext field that use DES \textit{grizY} band photometry are also magnitude-limited and have a completeness of 95 per\,cent to  24.6, 24.3, 24.0, 23.7. and 23.4\,mag respectively \citep{Abbott2021}. All magnitude limit values are for point sources.

After positionally cross-matching the WiggleZ-type objects with FLASH island catalogues, we obtain `associated' galaxies (<5\,arcsec separation), and `offset' galaxies (between 5 and 20\,arcsec separation). As we explore via BPT \citep{Baldwin1981} and WHAN \citep{Fernandes2010,Fernandes2011} spectral line ratio diagnostic diagrams, the mass-excitation \citep[MEx;][]{Juneau2014} diagnostic diagram, and the evolution of (\textit{g}\,--\,\textit{i}) colours over redshift in \citet{Eden2025}, these UV-selected associated and offset galaxies are generally categorised as different kinds of galaxies, where offset galaxies are >80 per\,cent SF, while associated galaxies contain 2\,--\,5 times the number of galaxies that are active galactic nuclei (AGN) compared to both the offset category, and WiggleZ-type galaxies as a whole. \citet{Jurek2013} find that the median SFR of WiggleZ galaxies varies from $\sim{1}$\,$\pm$\,0.5\,M$_{\odot}$\,$\mathrm{year^{-1}}$ at $z$\,=\,0.4, to $\sim{18}$\,$\pm$\,3\,M$_{\odot}$\,$\mathrm{year^{-1}}$ at $z$\,=\,1.0.

\subsection{The WiggleZ spare fibre sample}\label{spare_radio_sample}

In \citet{Eden2025}, we introduced the SPARE\_RADIO class of objects studied in the WiggleZ survey. These are Faint Images of the Radio Sky at Twenty-cm \citep[FIRST;][]{Becker1995} radio galaxies targeted in the Large Area Radio Galaxy Evolution Spectroscopic Survey \citep[LARGESS;][]{Ching2017}, where they searched for optical components of these radio galaxies.

LARGESS galaxies were observed by WiggleZ as part of its effort to allocate fibres towards companion projects whenever high priority WiggleZ were unable to be targeted \citep{Drinkwater2018}. WiggleZ, along with GAMA, provided optical observations and spectroscopic redshifts to LARGESS. Unlike the WiggleZ galaxies, these galaxies are not UV-selected, but GALEX MIS NUV and FUV values have been measured where possible.

These radio galaxies lie within the same redshift range as the WiggleZ galaxies, and essentially form a `control sample' of radio galaxies with which to compare the star-forming WiggleZ-type associated galaxies. The median redshifts of the associated, offset, and SPARE\_RADIO samples are 0.684, 0.625, and 0.593 respectively. Given the selection criteria for the WiggleZ galaxies is different than for the SPARE\_RADIO sample, the associated and offset samples possessing more similar median redshifts compared to the SPARE\_RADIO sample seems reasonable, although all are close to the WiggleZ median target redshift of $\sim{0.6}$. The median stellar masses for those galaxies that have values calculated in the associated, offset, and SPARE\_RADIO samples are 10.85, 10.40, and 11.20 log$_{10}$\,($\mathrm{M_{\star}/M_{\odot}}$) respectively. While all three values are close, as expected, the SPARE\_RADIO sample has the highest median stellar mass, followed by the star-forming associated sample which contains a large proportion of radio galaxies, followed by the star-forming offset sample.

Out of the total 398 SPARE\_RADIO galaxies that lie within the FLASH fields listed in Table \ref{tab:field_sbids}, we only carry out spectral investigations of the 23 SPARE\_RADIO galaxies possessing a FLASH component within 5\,arcsec, and with FLASH peak fluxes of 30\,mJy or greater. Throughout the rest of this work, we refer to these galaxies as the spare fibre sample.

\begin{table}
\centering
\begin{tabular}{l|l|l|l}
    \hline \hline
    Field   & FLASH field & \citet{Eden2025} & Updated FLASH \\
      &  & FLASH SBID & SBID \\
    \hline    
    W09 & 546  & 34599 & 55399 \\
        & 547  & 34549 & 59842 \\
    W15 & 559  & 41085 & 51446 \\
        & 560  & 34563 & 52528 \\
        & 561  & 34576 & 50021 \\
    Ext & 526  & 42275 & 55394 \\
        & 527  & 42300 & 42300 \\
        & 528  & 37450 & 51440 \\
        & 529  & 34557 & 51450 \\
        & 576  & 34556 & 34578 \\
        & 577  & 42296 & 62514 \\
        & 578  & 43424 & 62515 \\
        & 579  & 42298 & 42298 \\
        & 580  & 34567 & 62806 \\
        & 581  & 41225 & 62854 \\
        & 582  & 34578 & 51449 \\
    \hline
\end{tabular}
\caption[Table of full FLASH survey fields corresponding to W09, W15, and Ext fields]{The three major fields in this work, the unique SBIDs of the FLASH field catalogues covering parts of the field used in \citet{Eden2025}, and the most current FLASH field SBIDs used in this work. Due to the irregular distribution of objects in the extended field, three of the FLASH fields that cover the contiguous area of the extended field did not contain a FLASH source in sufficient proximity to match against a WiggleZ-type galaxy or spare fibre galaxy, and thus are not included in the table. These are fields 525, 530, and 531.}
\label{tab:field_sbids}
\end{table}

\begin{figure*}
    \centering
    \includegraphics[width=\linewidth]{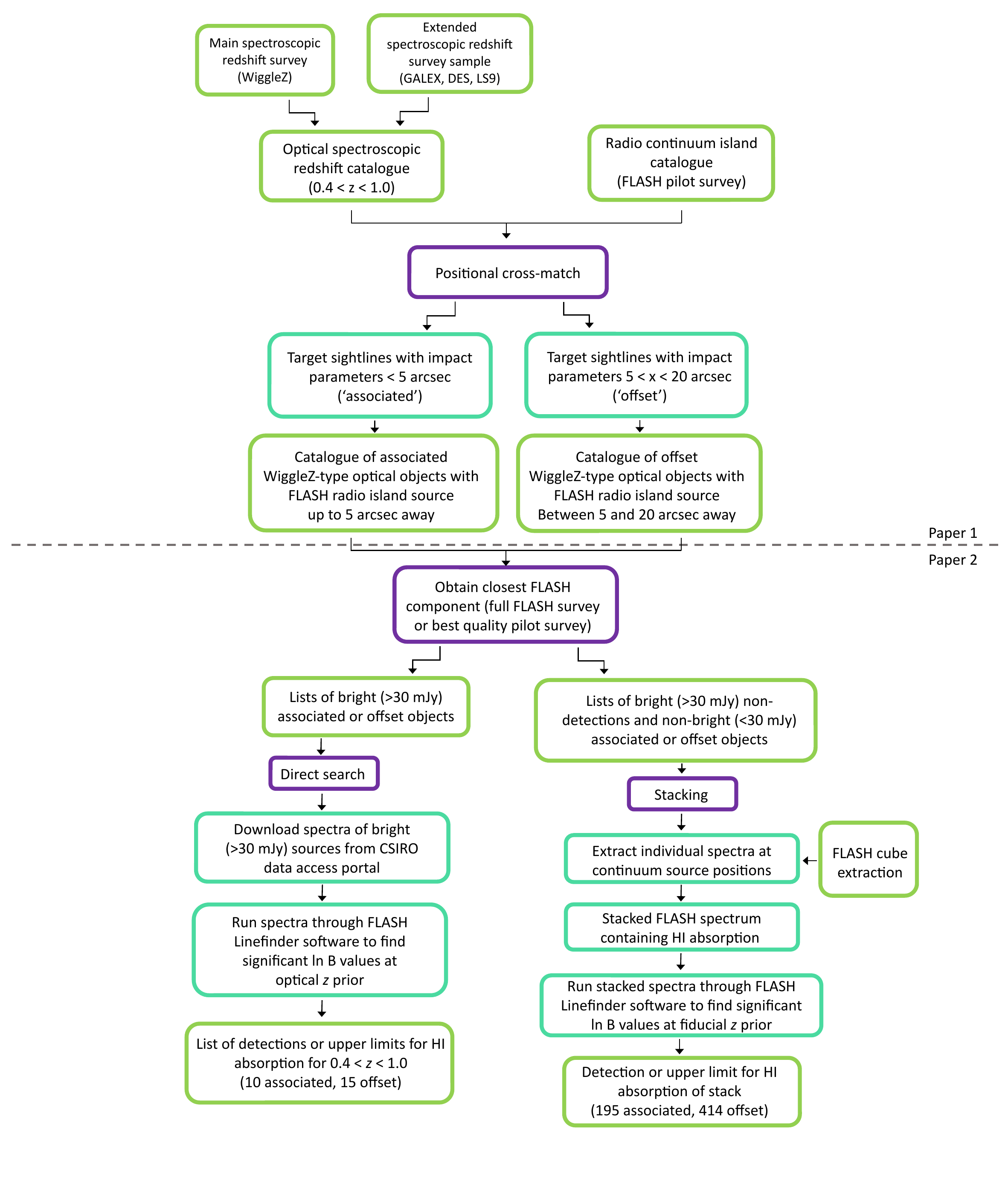}
    \caption[Flowchart of extended WiggleZ-type galaxy sample production]{Production process of the associated and offset object samples investigated in this work, as well as outline of the two kinds of \hi 21\,cm absorption search (direct and stacking) which are carried out. The scope of work covered in \citet{Eden2025} and this work are separated by the grey horizontal dashed line.}
    \label{fig:flowchart}
\end{figure*}

\section{FLASH radio spectra}\label{flash_sources}

\subsection{FLASH spectra}\label{flash_spectra}

The First Large Absorption Survey in \hi (FLASH) is a currently in-progress ASKAP survey designed to detect \hi 21\,cm absorption across the redshift range 0.42\,<\,\textit{z}\,<\,1.0 \citep{Allison2022,Yoon2025}. FLASH has a spatial resolution of 30\,arcsec and spectral resolution of 18.5\,kHz \citep{Yoon2025}. At time of writing, the full survey is currently around 30 per\,cent complete. The FLASH survey area covers much of the celestial equator. FLASH pilot survey island catalogues were used in \citet{Eden2025} to produce the resulting catalogue of optical-radio cross-matched objects. In this work, we introduce component catalogues from the full FLASH survey, obtaining the FLASH component nearest the optical WiggleZ-type object from the cross-matched FLASH island. The \textsc{selavy} \citep{Whiting2012} source-finding algorithm is used as part of the \textsc{askapsoft} \citep{Whiting2020} pipeline to calculate the likelihoods of association between component sources in the 3D sky, and flags whatever source appears to be at the group's centre. The FLASH component is the radio source we use throughout this work for investigating associated and stacked absorption.

We ensure that the optical galaxies we are probing have been cross-matched against the `a' component of the FLASH source. Simple FLASH sources will always be `a' components. When a FLASH source has been found to be `complex' by \textsc{selavy}, its components are named in order of decreasing brightness, with `a' awarded to the brightest, `b' to the second brightest, and so on. We are only interested in matches against primary sources, which are either the sole component of a simple source, or the component most likely to be the bright central component of a complex source, so we limit to `a' components. In \citet{Eden2025}, we found that only $\sim{}$10\,per\,cent of FLASH islands cross-matched with optical galaxies were complex and contained more than one component, a very similar result to the rates of simple and complex matches between optical SDSS galaxies and FIRST radio sources found by \citet{Ivezic2002}.\color{black}

Some FLASH fields have repeat observations. Each time a field is observed, the new observation is designated with a new ASKAP scheduling block ID (SBID). Each SBID is unique, and allows different observations of the same field to be distinguished. SBIDs are contained in all FLASH data products, including island and component IDs. The catalogues of associated and offset cross-matches in \citet{Eden2025} contained a column of FLASH island IDs for each radio source, and the SBIDs within these names were from Pilot Survey observations.

The FLASH Pilot Survey continuum data used to make the catalogues in \citet{Eden2025} are of good quality, but some Pilot Survey spectral line data contains artefacts \citep{Yoon2025}, which affect spectral data quality. We now use observations from the full FLASH survey for the analysis in this paper which generally are of higher quality or less affected by artefacts. As such, the SBIDs used in \citet{Eden2025} will in most cases have been replaced in this work by full survey SBIDs (see Table\,\ref{tab:field_sbids}).

There are exceptions to this, such as in fields 525, 527, 576, and 579, which retain SBIDs beginning with either 3 or 4, instead of 5 or 6, indicating their pilot survey status. We use them now as these SBIDs currently have the best quality observation for the field. This can be due to a variety of reasons, such as later observations having been noted as being more severely affected by atmospheric ducting of RFI, most commonly at 758\,--\,788\,MHz, 870\,--\,890\,MHz, and 943\,--\,960\,MHz (see \citet{Yoon2025} for more detailed discussion of RFI and its impacts on FLASH). These issues are flagged in unique validation reports produced for each SBID, noting issues with catalogues, images, or cubes, allowing users of the data products to be aware of any issues with data that may adversely affect any research being attempted. For the FLASH spectra, the typical root mean squared (rms) spectral noise is between 4 -- 5 mJy\,bm$^{-1}$ \citep{Yoon2025}.

The general procedure for work in this series is shown by the flowchart in Fig.\,\ref{fig:flowchart}. The production of the associated and offset catalogues is shown in the top region of this figure (labelled `Paper 1'), outlining in brief how these catalogues were made. In the subsequent `Paper 2' region, the processes of obtaining the final results of this paper for this series are outlined. We obtain the closest FLASH component for each associated and offset object from either the full FLASH survey, or whichever pilot survey SBID has been validated as the best quality observation for this field. This will allow us to obtain the best quality spectra possible for analysis.

Bright sources (with peak flux density >\,30\,mJy) are our targets for direct searches for 21\,cm absorption, in the associated regime probing gas within the galaxy, and the intervening regime probing gas in the CGM. \citet{Banerji2013} find stellar masses of $10^{9.5}$\,--\,$10^{11}$\,M$_{\odot}$ for 40,000 WiggleZ galaxies at 0.3\,<\,$z$\,<\,1.0. \citet{Wisnioski2011} study star-forming WiggleZ galaxies at $z \sim{1.3}$ and find they have H$\upalpha$ `half-light' radii ($r_{1/2}$) of 1.8\,--\,4.2\,kpc, while \citet{Ching2017} shows that the LARGESS galaxies comprising our sample of spare fibre radio galaxies have $r$-band half-light radii of $\sim{8}$\,kpc for $z$\,>\,0.3. \citet{Somerville2018} outlines the relationship between $r_{1/2}$ and the virial radii of galactic dark matter halos ($r_{\mathrm{vir}}$) as $r_{1/2}$ = 0.018\,$r_{\mathrm{vir}}$, and which shows weak or negligible redshift evolution for the interval 0.1\,<\,$z$\,<\,3.0. These half-light radii then equate to virial radii of $\sim{100}$\,--\,233\,kpc for WiggleZ galaxies, and $\sim{444}$\,kpc for spare fibre galaxies.

At the median WiggleZ redshift of z\,=\,0.6, the angular scale on the sky is 6.76\,kpc\,arcsec$^{-1}$. Thus, a 5\,--\,20\,arcsec offset would correspond with an impact parameter of around 30\,--\,120\,kpc and would typically probe the CGM beyond the galaxy disc but within the virial radius, while an offset smaller than 5\,arcsec mainly probes gas within the galaxies themselves. Any detections from these bright associated and offset objects will thus tell us about two different kinds of environment. As the spare fibre galaxies have 5\,arcsec separations, searches around these galaxies as with the associated galaxies probes the gas within the galaxies themselves.

Any sources which do not meet the criteria for classification as bright FLASH sources (<\,30\,mJy) are referred to as `non-bright'. Unlike the bright source spectra which are available on CSIRO ASKAP Science Data Archive (CASDA; \url{https://research.csiro.au/casda/}), non-bright spectra must be manually extracted from FLASH cubes. These spectra are obtained by producing cutouts from the FLASH spectral cubes at the positions of the FLASH sources. For the three groups of galaxies (associated, offset, spare fibre), we stack the spectra, producing co-added spectra for each different group of cross-matched galaxies.

FLASH spectral cubes undergo post-processing in the \textsc{askapsoft} pipeline to remove the continuum. Continuum subtraction is first done in the visibility domain, and then by subtracting the image-based continuum across 5\,MHz intervals rather than smaller intervals like 1\,MHz, in order to avoid completely subtracting out \hi 21\,cm lines with widths of 300\,km\,s$^{-1}$. Further continuum subtraction is done via the fitting of a second-order polynomial. Full detail on the continuum subtraction process for the FLASH survey, including issues encountered in the pilot surveys, is given in \citet{Yoon2025}.

Additionally, we can evaluate the effect of all sources of error on FLASH observations - including uncertainties in continuum subtraction - by estimating the reproducibility of \hi line measurements using the \textsc{flashfinder} tool (full detail on \textsc{flashfinder} is given in Section\,\ref{direct_search}). \citet{Yoon2025} show the reproducibility of \textsc{flashfinder} \hi 21\,cm absorption detections by comparing the measurements of redshift ($z$), peak optical depth ($\tau_{\mathrm{peak}}$), integrated optical depth ($\tau_{\mathrm{int}}$), and velocity width ($\upDelta v$) returned for lines detected in FLASH Pilot 1 or Pilot 2, and re-observed during Pilot 2. They find that redshift measurements are highly reproducible (differing by 0.1\,per\,cent). The peak and integrated optical depth measurements have typical uncertainties of around 25\,per\,cent, and the line width measurements (fitting a single component) have a typical uncertainty of around 20\,per\,cent, though with a small number of outliers in each case. This shows that the \textsc{flashfinder} measurements are quite reproducible, meaning that uncertainties in continuum subtraction as part of all sources of error have only a small effect on detections. The effect of this on weak or broad detection lines should in general not be too large.

\subsection{Optical depth and column density}\label{optical_depth}

For scientific conclusions to be drawn from the FLASH spectra, conversion from flux density into units of optical depth is needed. Eq. \ref{eq:optical_depth} shows the relationship between optical depth ($\tau$), (continuum) flux density ($S_{\mathrm{c}}$), spectral line depth ($\Delta S$), and covering factor ($f_{\mathrm{c}}$) in terms of velocity ($\nu$):

\begin{equation}
    \tau(\nu) = -\ln\,\Bigl( 1 - \frac{\Delta S(\nu)}{f_{\mathrm{c}} S_{\mathrm{c}}} \Bigr) \label{eq:optical_depth}
\end{equation}

\smallskip
\noindent
The covering factor accounts for the fraction of the radio source flux intercepted by the absorber. Since the vast majority of 21\,cm absorbers are optically thin \citep{Curran2013a}, then $\Delta$S/$f_{\mathrm{c}} S_{\mathrm{c}}$ $\lesssim$ 0.3 and thus eq. \ref{eq:optical_depth} would not converge. Therefore we can say that:

\begin{equation}
    \tau(\nu) \approx \frac{\Delta S(\nu)}{f_{\mathrm{c}} S_{\mathrm{c}}} \approx \frac{\tau_{\mathrm{obs}}(\nu)}{f_{\mathrm{c}}} \label{eq:assumption}
\end{equation}

\smallskip
\noindent
In the optically thin regime, the column density of the gas (atoms\,cm$^{-2}$) is related to the integrated optical depth of the 21\,cm absorption via:

\begin{equation}
\begin{split}
    N_{\hi} &\approx  1.823 \times 10^{18}\,T_{\mathrm{s}} \int \tau(\nu)\,\mathrm{d}\nu\\
    &\approx  1.823 \times 10^{18}\,\frac{T_{\mathrm{s}}}{f_{\mathrm{c}}} \int \tau_{\mathrm{obs}}(\nu)\,\mathrm{d}\nu\\
    \label{eq:column density}
\end{split}
\end{equation}

\smallskip
\noindent
Where $T_{\mathrm{s}}$ is the spin temperature in K, and velocity is in km\,s$^{-1}$. Thus, high integrated optical depths are indicative of low spin temperatures, or high covering factors. For further detail on these relationships, see \citet{Rohlfs2004}.


\section{Direct absorption search}\label{direct_detections}

\subsection{Direct detections}\label{direct_search}

\begin{figure}
    \centering
    \includegraphics[width=\linewidth]{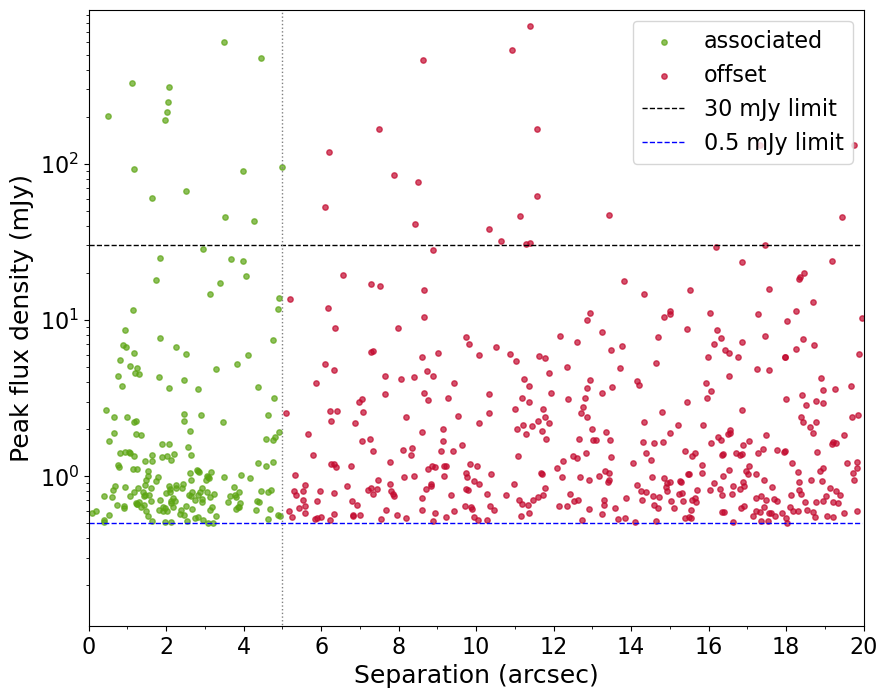}
    \caption[Peak flux density distribution of associated and offset data]{Peak flux density of each source in mJy against separation in arcsec between the optical object and the FLASH radio source for the associated (green) and offset (pink) objects, for 0.42\,<\,\textit{z}\,<\,1.0. The horizontal black dashed line at 30\,mJy marks the distinction for `bright' objects, where anything above 30\,mJy is considered bright. The 5$\upsigma$ limit for FLASH source detections is peak flux density of 0.5\,mJy. None of the cross-matched objects in this sample are below that limit, due to cuts put in place that are outlined in section 3.2 of \citet{Eden2025}. Objects below the vertical dotted grey line at 5\,arcsec are associated, and above it are offset. There are 194 non-bright associated objects and 15 bright associated objects, and 433 non-bright offset objects and 21 bright offset objects.}
    \label{fig:peak_flux_sep}
\end{figure}

For direct \hi 21\,cm absorption searches to be successful, they are dependent on a background radio source to be bright enough for a detection. Alternatively, in the case no detection can be made, a radio source needs to be bright enough to put in place a restrictive optical depth limit. In our case, as we are targeting observations towards galaxies, accurate redshifts are used as priors to search for absorption signals around, which are provided by spectroscopic redshifts. The typical WiggleZ redshift uncertainty is 48\,km\,s$^{-1}$ or better for redshifts with higher quality flags \citep{Drinkwater2018}.\color{black}

For FLASH sources, we are interested in anything exceeding 30\,mJy peak flux density cross-matched with a galaxy at 0.42\,<\,\textit{z}\,<\,1.0. In Fig.\,\ref{fig:peak_flux_sep} both the populations of associated and offset cross-matched objects have their flux density plotted against the separation between optical WiggleZ-type object and the FLASH radio source. It can be seen that the majority of both associated objects (194/209; 92.8\,per\,cent) and offset objects (433/454; 95.4\,per\,cent) fall beneath the 30\,mJy flux density cut.

The remaining 15 associated and 21 offset objects are bright enough (peak flux density >\,30\,mJy) to offer the best chances of \hi absorption detections. After removing objects cross-matched against non-primary component FLASH sources, there are then 10 bright associated objects, and 15 bright offset objects in total. Their FLASH component spectra are available from CASDA. Redshift priors are taken from the spectroscopic redshift of the optical WiggleZ-type object. A search for lines within the spectra is then done using the \textsc{flashfinder} software which was developed to find lines in FLASH spectra \citep{Allison2012}. This uses Bayesian statistics to find possible absorption features in a spectrum, and returns the redshift of the possible feature and its associated Bayesian evidence value (ln\,B), which indicates the statistical significance of a detection, reflecting the number of sigma from a null result - in this case, no line existing in the spectrum - that the detection is \citep{Trotta2008,Yoon2025}.

In a blind search, ln\,B values above 10 can indicate a detection, where ln\,B\,=\,10 would require a spectrum free of artefacts for a detection to be able to be considered astronomical in nature, and not a result of noise or artefacts, while ln\,B\,=\,30 would indicate a highly significant detection \citep{Yoon2025}. For reference, ln\,B\,=\,5 would correspond with a statistical significance of 3.6$\upsigma$, indicating a detection that is `strong (at best)' \citep{Trotta2008}, while ln\,B\,=\,11 would correspond with a 5$\upsigma$ detection \citep{Trotta2008,Yoon2025}, though it should be noted that this assumes that noise and artefacts are Gaussian, which is known not to be the case for FLASH spectra \citep{Yoon2025}, hence the higher requirement of ln\,B\,$\geq{30}$ for blind FLASH detections to be considered significant.

However, for detections with a redshift prior, confidence can be much higher, with ln\,B values as low as 8 indicating a detection. \textsc{flashfinder} is used to search our bright FLASH component spectra for \hi 21\,cm absorption detections at or around the spectroscopic redshifts of the optical WiggleZ galaxies. \color{black}


\subsection{Bright galaxy search results}\label{bright_search_results}

After running \textsc{flashfinder} to search for significant detections at or around the optical redshift priors for WiggleZ-type galaxies, we do not find any significant detections for the 10 associated and 15 offset galaxies. Upper limits for observed optical depths, integrated optical depths, and column densities have been determined (see Table \ref{tab:bright_object_values}).

As discussed in \citet{Eden2025} and within this work, we expected that our sample of UV selected SF galaxies would be excellent candidates in a search for \hi 21\, absorption, inferring that \hi discs would be able to be detected for both associated and offset objects with sufficiently bright radio sources in their vicinities due to the galaxies possessing \hi reservoirs that would be fuelling their ongoing star formation. We do not see this effect, and we discuss possible causes of this lack of detection in detail in Section\,\ref{limiting_factors}.

\subsection{Spare fibre search results}\label{spare_radio_results}

We find 23 spare fibre galaxies that fit our desired criteria which is matched with a bright primary FLASH component. Using optical redshift priors, we find only one \hi 21\,cm absorption detection which is towards the radio galaxy NVSS J214954-004657, at \textit{z}$_\mathrm{opt}$ = 0.864. This gives a detection fraction of $4.3^{+1.9}_{-1.4}$\,per\,cent where errors are 1$\upsigma$ Gaussian errors taking into account small number statistics \citep{Gehrels1986}.

This detection rate is consistent with the rate of detection in similar searches at 0.42\,<\,\textit{z}\,<\,1.0 with FLASH. \citet{Su2022} searched for associated absorption in GAMA galaxies, and found a detection rate of $2.9^{+9.7}_{-2.6}$\,per\,cent, while \citet{Aditya2024} targeted 1-Jy Molonglo Reference Catalogue (MRC) galaxies, and found a detection rate of $1.8^{+4.0}_{-1.5}$\,per\,cent.

All remaining galaxies in the spare fibre sample do not possess a significant \hi 21\,cm detection towards the spectroscopic redshift of the target galaxy. Upper limits for observed optical depths, integrated optical depths, and column densities for these galaxies are included in Table \ref{tab:spare_radio_object_values}.

\subsubsection{Object NVSS J214954-004657}\label{direct_detection_spare_radio}

In the FLASH spectrum of radio galaxy NVSS J214954-004657 we find a \hi 21\,cm absorption feature at \textit{z}$_{\mathrm{HI}}$ = 0.863, slightly blueshifted from the galaxy redshift of \textit{z}$_{\mathrm{opt}}$ = 0.864, corresponding with a velocity difference of $\approx{-160}$ km\,s$^{-1}$. \textsc{flashfinder} determined a ln\,B value of 8.0, which is statistically significant with the redshift prior.

The detection has a peak observed optical depth of 0.066\,$\pm$\,0.014, and an integrated optical depth of 6.27\,$\pm$\,1.33 km\,s$^{-1}$, implying that the detection has a significance of 4.7$\upsigma$. The width of the absorption line feature is 95\,km\,s$^{-1}$, which is typical for associated \hi 21\,cm absorption lines detected with FLASH \citep[see fig.\,20 of][]{Yoon2025}. This radio galaxy was first targeted in the LARGESS survey, and its optical component was first observed by WiggleZ. The corresponding FLASH component is 62514\_component\_52a, and this detection has successfully been re-observed using \textsc{flashfinder} in 42296\_component\_52a and 34555\_component\_53a, with consistent parameters (see Table\,\ref{tab:spare_fibre_linefinder}).

\begin{figure}
    \centering
    \includegraphics[width=\linewidth]{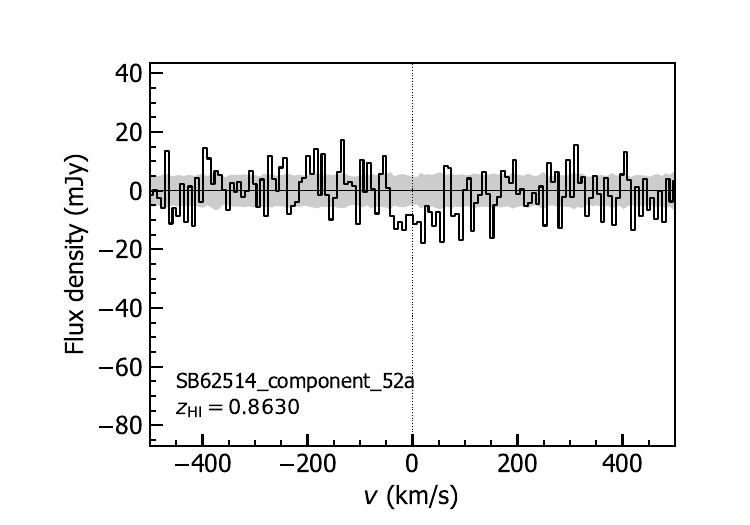}
    \caption[Detection towards NVSS J214954-004657]{Absorption line detected in spare fibre galaxy NVSS J214954-004657. The feature was detected close to the optical redshift of the galaxy (\textit{z}$_{\mathrm{opt}}$ = 0.864). The ln\,B value for this detection is 8.0, which is statistically significant with the redshift prior. The grey region is one times the rms spectral-line noise.}
    \label{fig:spare_radio_detection}
\end{figure}

\section{Stacking}\label{stacking}


\subsection{Stacking methodology}\label{stacking_methodology}

Spectral stacking is an analytical technique that involves averaging over many different spectra in order to observe faint emission/absorption lines which otherwise may lie in spectra undetected, due to low signal-to-noise ratios. It requires the co-addition of a large number of spectra significantly reducing the noise level, while preserving any spectral line signal.

As we are considering spectra over a large range of redshifts, they need to be rebinned according to redshift to account for the relative broadening and narrowing of lines of potential hidden \hi 21\,cm absorption signals at different redshifts due to spectral channel width being fixed in frequency. We use a \textsc{python} module named \textsc{linestacker} \citep{Jolly2020} to stack our spectra. \textsc{linestacker} allows for the rebinning of spectra according to either the minimum redshift in the stack, or the maximum. The option for minimum was the default, and we continued with this option.

\textsc{linestacker} rebins the spectra in the stack and shifts them into the rest-frame frequency according to the spectroscopic redshift. The frequency axis of the stacked spectra are then converted into velocities with respect to the \hi 21\,cm rest frequency (1420\,MHz). If a stacked signal exists, it would be expected to be seen at or around the rest velocity of 0\,km\,s$^{-1}$.

While \hi 21\,cm emission line detection is performed in terms of flux density, \hi 21\,cm absorption detections must be carried out in terms of optical depth (opd) for physical conclusions to be made. This carries across to stacking experiments, where emission stacking is done in terms of flux density, but absorption spectra must be converted into opd.

To test that noise in the stacked spectra produced by \textsc{linestacker} follow expected behaviour, we follow emission stacking procedure and stack in terms of spectral flux density. As long as the noise within the spectra is Gaussian, rms noise is expected to fall as $N^{-0.5}$, where $N$ is the number of spectra being co-added into the stack \citep{Delhaize2013,Healy2019,Veronese2025}. In order to be less susceptible to outliers due to a single bad channel or large intensity, we choose to take the median channel values of the $N$ spectra in the stack, rather than the mean.

\begin{figure}
    \centering
    \includegraphics[width=\linewidth]{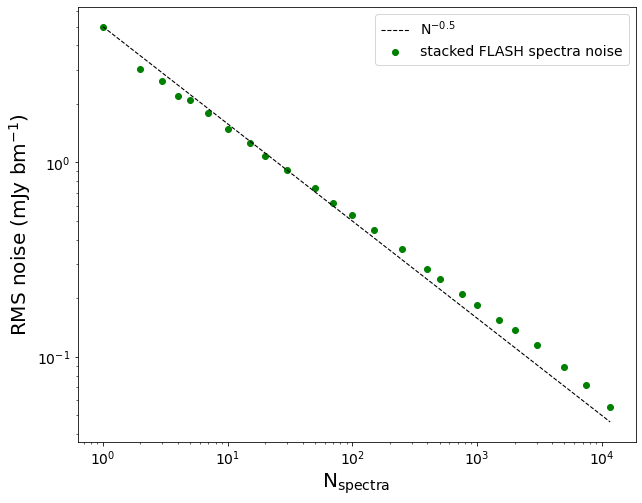}
    \caption[Stack rms noise value versus number of stacked spectra]{Stack rms noise value (mJy\,bm$^{-1}$) against number of spectra ($N_{\mathrm{spectra}}$) for all 11653 sources with brightness <\,30\,mJy\,bm$^{-1}$ found in FLASH field 559, SBID\,51446 (green points). This cube was validated as `GOOD' quality and lacks spurious spectral lines due to atmospheric ducting and RFI. A line of $N^{-0.5}$ is plotted (dashed black line).}
    \label{fig:stacking_vs_noise}
\end{figure}

Fig.\,\ref{fig:stacking_vs_noise} shows the relationship between number of spectra being stacked, and the rms noise of the stacked spectra (green points) from FLASH SBID 51446. The expected behaviour of noise decreasing as $N^{-0.5}$ is shown with the dotted black line. The departure of the data from decreasing as $N^{-0.5}$ at $N\sim{300}$, is seen by \citet{Fabello2011} with their \hi 21\,cm emission stacking as a result of non-Gaussian noise beginning to become dominant in the stack. Non-Gaussian noise and artefacts have been a known issue in FLASH pilot survey data \citep{Yoon2025}.

The validation reports produced for each SBID were used to identify particular frequency regions affected by ducting or interference. This allowed for masking of problematic regions of spectra according to the SBID of the observation. Masking these regions ensures that spurious signals do not contribute to potential stacking detections, create the appearance of a detection where there is none, or otherwise interfere with a genuine detection.\color{black}

As CASDA only hosts bright FLASH spectra, the spectra for offset stacking require extraction directly from spectral cubes. We developed a pipeline that allowed extraction of FLASH cube cutouts in a 3\,x\,3 pixel cube centred on the position of the FLASH source, corresponding to a 6\,arcsec radius cutout, for the full FLASH frequency range of 711.5\,MHz -- 999.5\,MHz. These cubes were then converted into 1\,dimensional spectra by taking median values across the 9 spectra contained in each cube cutout, in order to mitigate any outliers or bad pixels.

Like the FLASH cubes, the spectral flux density axes of these cutouts is in units of Jy\,bm$^{-1}$, and require conversion into opd for conclusions to be made. The offset spectra are converted into mJy\,bm$^{-1}$, before being normalised against the integrated flux of the source, resulting in units of $\Delta S$/$S_{\mathrm{cont}}$.

\subsection{Stacking results}\label{stacking_results}

Initially, we attempted three separate stacks for the associated, offset, and spare radio samples by stacking all available spectra randomly, in no particular order. As in Section\,\ref{direct_search}, any cross-matches against complex (not `a') sources are removed, dropping non-bright associated objects from 194 to 185, and non-bright offset objects from 433 to 400.  As we are looking for any hidden signal within the stacked spectra, the inclusion of confirmed or potential detections could contaminate the signal. For this reason, the spectrum of 62514\_component\_52was removed from the spare fibre stack, meaning that 10 bright associated objects, 14 bright offset objects, and 22 bright spare fibre objects will be stacked. This leaves 195, 414, and 22 spectra for the associated, offset, and spare fibre stacks respectively.

Stacking in this way, we did not find a 21\,cm absorption feature for any of the three stacks. As an independent check, the stacking was also performed with \textsc{`\hi Stacking Software'} \citep[\textsc{hiss;}][]{Healy2019}. This is a spectral stacking software developed in \textsc{python} specifically to analyse radio spectra that are thought to contain a \hi 21\,cm emission or absorption line. Stacking using \textsc{hiss} similarly does not yield a \hi 21\,cm absorption feature from within the stacks. It also shows that the rms noise of our stacked data declines as $N^{-0.5}$ until $N\,\sim{300}$, similarly to Fig.\ref{fig:stacking_vs_noise}, reconfirming that non-Gaussian noise is becoming dominant in FLASH spectra beyond $N\sim{300}$.

\begin{figure}
    \centering
    \includegraphics[width=\linewidth]{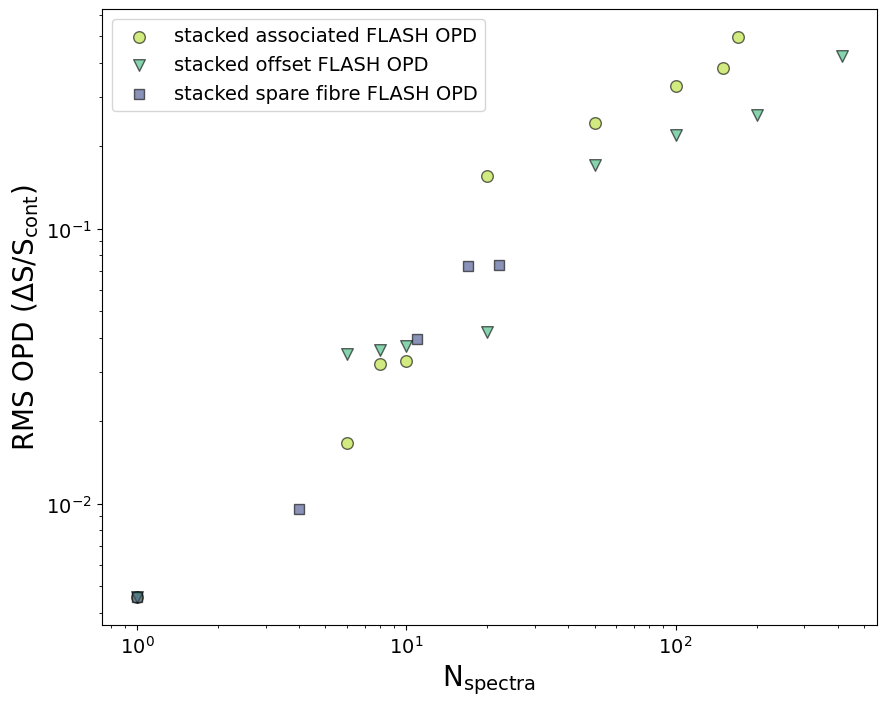}
    \caption[Stack rms opd values for the associated, offset, and spare fibre samples arranged in order of descending brightness]{Stack rms opd values ($\Delta$S/S$_\mathrm{cont}$) against number of spectra ($N_{\mathrm{spectra}}$) for all 195 associated, 414 offset, and 22 spare fibre sources that are bright (>\,30\,mJy) containing non-detections, or non-bright (<\,30\,mJy). The spectra being stacked are stacked in order of decreasing brightness. Associated stacks are plotted with green circles, offset with teal triangles, and spare fibre with purple squares. The increase of rms opd in stacks with increasing $N_{\mathrm{spectra}}$ is clear in all three samples.}
    \label{fig:rms_opd_vs_n}
\end{figure}

Unlike previous stacking experiments, we then investigate the effect of stacking order. We stack according to the peak flux density of each object in descending order, instead of randomly, stacking from brightest to least bright. The effect of this ordered approach to the stacking can be seen in Fig. \ref{fig:rms_opd_vs_n}. The opd of the stacked spectra increase as a greater number of stacked spectra are co-added in this order.

This is due to the nature of the relationship between bright radio sources and optical depth. Where bright sources have low optical depth, in the course of co-adding an increasing number of less bright objects into a stack, the optical depth will gradually be brought up as the median optical depth value of the stacked spectra increases with the number of objects.

When we visually inspect the stacks between $N$\,=\,5 and the maximum number of spectra available for stacking, we see a potential recurring small signal at zero of the stacked spectrum rest frame. At a certain point in the set of optical depth stacks, this potential signal is no longer apparent. It is possible that a \hi 21\,cm absorption signal does lie buried within the stacked data, but that above a certain optical depth this is no longer able to be detected. We use \textsc{flashfinder} to investigate the associated, offset, and spare fibre stacks to look for evidence of this signal.

\subsubsection{Associated stacking search}\label{associated_stacking}

\begin{figure}\includegraphics[width=\linewidth]{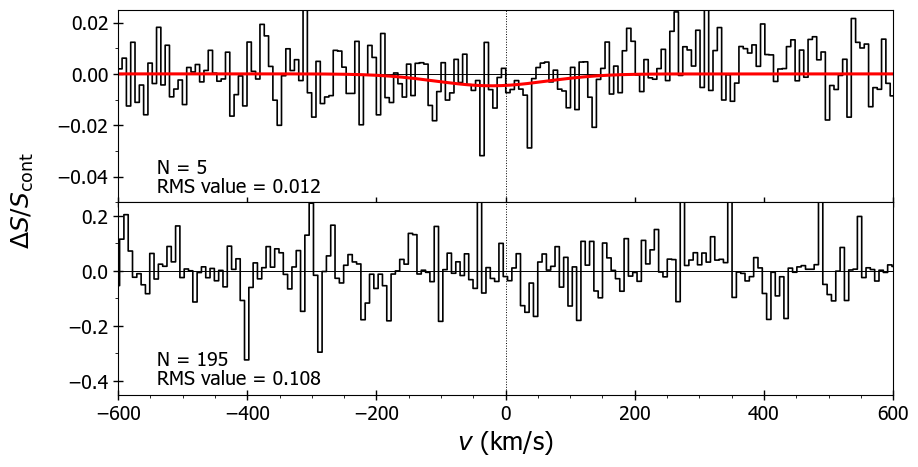}

\caption[Stacked associated optical depth]{Stacked spectra produced by stacking first 5 and then all 195 bright and non-bright associated FLASH spectra in terms of optical depth in order of decreasing brightness. The \textsc{flashfinder} detection for the $N$\,=\,5 stack is indicated in red. Values of rms opd are included for each stack. The opd rms values increase to a maximum value when all spectra have been stacked. Note that the vertical scale changes to best display the data.}
\label{fig:associated_stack}
\end{figure}

In the case of the associated stacks, \textsc{flashfinder} flags a tentative \hi 21\,cm absorption detection for the $N$\,=\,5 stack of Fig.\,\ref{fig:associated_stack}, very close to the central redshift, with a velocity difference of -17.5 km\,s$^{-1}$. The peak observed optical depth is 0.006\,$\pm$\,0.001, the observed velocity integrated optical depth is 1.280\,$\pm$\,0.156 km\,s$^{-1}$, and the line width is 227\,km\,s$^{-1}$. The total flux of the co-added radio sources in this stack is 1497.79\,mJy.  The ln\,B value assigned by \textsc{flashfinder} is 36.9, indicating high statistical significance. However, this large ln\,B value could be high due to the broadness of the fitted line, and this kind of broad feature in the stack could arise from various factors, such as imperfect continuum subtraction in the five spectra in this stack. For this reason, we describe this stacked feature as tentative.

\textsc{flashfinder} does not flag any additional detections with meaningful ln\,B values above 8 for any of the remaining associated stacked spectra. It is reasonable to assume that the stacked signal, if astronomical in nature, begins to become buried in this range as optical depth increases. The total flux of the co-added radio sources is 2454.52\,mJy for $N$\,=\,195.

We again take care to acknowledge that a detection with a high statistical significance does not automatically imply that the detection is astronomical in nature - merely that the detection is significant in comparison to the noise fluctuations around it. Additionally, the behaviour of stacked spectra is different and more complex than that of individual spectra, and stacked absorption spectra in terms of optical depth even more so. Where one can use an optical redshift and other knowledge of a galaxy for higher confidences in \hi 21\,cm absorption searches with individual spectra, the same is not true with stacked spectra. Future work investigating the behaviour of \hi 21\,cm absorption stacking and the effect of the brightness of background sources will be needed.

\subsubsection{Offset stacking search}\label{offset_stacking}

\begin{figure}\includegraphics[width=\linewidth]{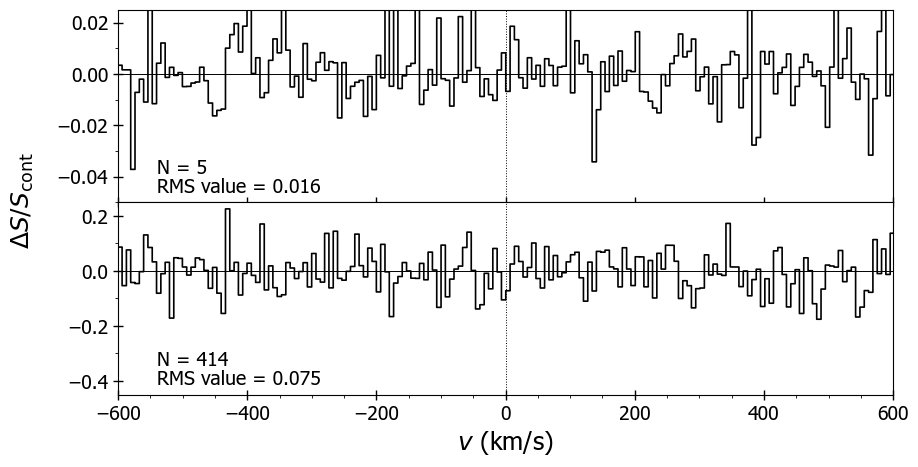}
\caption[Stacked offset optical depth]{Stacked spectra produced by stacking the first 5 and then all 414 bright and non-bright offset FLASH spectra in terms of optical depth in order of decreasing brightness. Values of rms opd are included for each stack. The opd rms values increase to a maximum value when all spectra have been stacked. Note that the vertical scale changes to best display the data.}
\label{fig:offset_stack}
\end{figure}

Interestingly for the offset stacks (Fig.\,\ref{fig:offset_stack}), \textsc{flashfinder} does not find a significant detection for any of the stacked spectra. The total flux of the background radio sources for $N$\,=\,5 is 1550.88\,mJy, while the total for $N$\,=\,414 is 3129\,mJy.

Although the total co-added background flux for the $N$\,=\,5 offset stack is similar to the $N$\,=\,5 associated stack, where a tentative feature was flagged, the offset stack does not contain one. There are a number of reasons we consider for why this may be the case. Firstly, we know that the geometry of the offset sample is different in comparison to the associated and spare fibre samples. While we know that the objects in the associated and spare fibre samples host the FLASH radio source, the offset sample does not. The lack of detection where we expect to find one could be due to geometric issues, such as the covering factor of the FLASH source being less than 100 per\,cent, or orientation effects due to the orientation of the offset galaxy being edge-on rather than face-on, something that is known to affect intervening \hi 21\,cm absorption detections \citep{Curran2016a}.

A further difficulty with intervening \hi 21\,cm absorption detections is highlighted by \citet{Reeves2016}, where at redshifts low enough for \hi 21\,cm emission detections to be made in all 17 of the galaxies they study (\textit{z} < 0.04), intervening absorption is rare to detect (4.3 per\,cent). They hypothesise that this is likely due to the cooler ($T_{\mathrm{s}}\lesssim{100}$\,K) \hi that produces absorption existing in isolated, compact clouds within the \hi discs of galaxies. This introduces an additional geometric complexity compared to associated absorption, reducing the likelihood of intervening absorption being possible. This effect will increase at higher redshifts, also contributing to lower covering factors.

If a significant 21\,cm detection is found at a similar number of stacked spectra and total background flux for the spare fibre stacks, then it would indicate that the effect is likely due to the nature of the offset galaxies as intervening \hi 21\,cm absorption targets rather than associated \hi 21\,cm targets. Alternatively, the degree of ionisation of the gas within the offset and spare fibre galaxies may need to be considered, if we see a similar lack of detections in the spare fibre sample.


\newgeometry{margin=1.5cm}
\begin{landscape}
\begin{table}
\renewcommand{\arraystretch}{0.8}
\centering
\begin{tabular}{c|c|c|c|c|l|c|c|r|c|c|c|c}
    \hline \hline
   no. & Galaxy name & \textit{z}$_{\mathrm{opt}}$ & FLASH & FLASH & match & $\upsigma_{\mathrm{\mathrm{rms}}}$ & $S_{\mathrm{c}}$ & $\Delta S_{\mathrm{peak}}$ & $\tau_{\mathrm{obs}}$ & $\int \tau_{\mathrm{obs}}\mathrm{d\nu}$ & $\log_{10}\biggl(\frac{N_{\mathrm{HI}} f}{T_{\mathrm{s}}} \biggr)$ \\
   &  &  &  SBID & component ID & type & (mJy\,bm$^{-1}$\,ch$^{-1}$) & (mJy) & (mJy) & & (km\,s$^{-1}$) & $\log_{10}[\mathrm{cm^{-2}\,K^{-1}}]$ \\
    \hline    
1  & J214712-012757 & 0.681 & 62514 & component\_295a & A & 4.90 & 42.8  & <1.4  & <0.033                 & <1.0  & <18.3 \\
2  & J212159-002612 & 0.885 & 34578 & component\_171a & A & 5.50 & 60.2  & <7.1  & <0.12                 & <3.5  & <18.8 \\
3  & J003727-001911 & 0.889 & 55394 & component\_33a  & A & 5.50 & 250.6 & <6.1  & <0.024                 & <0.73  & <18.1 \\
4  & J143833-000525 & 0.871 & 52528 & component\_11a  & A & 5.03 & 602.8 & <0.21  & <0.35$\times 10^{-3}$ & <0.010  & <16.3 \\
5  & J013451+000212 & 0.575 & 51440 & component\_26a  & A & 4.80 & 312.9 & <3.9  & <0.012                 & <0.37  & <17.8 \\
6  & J140558+000329 & 0.596 & 51446 & component\_160a & A & 5.18 & 66.9  & <5.1  & <0.076                 & <2.3  & <18.6 \\
7  & J091359+001315 & 0.755 & 59842 & component\_276a & A & 4.47 & 45.9  & <3.3  & <0.072                 & <2.2  & <18.6 \\
8  & J221029+010843 & 0.744 & 62515 & component\_30a  & A & 4.72 & 331.4 & <1.4  & <0.004                 & <0.13  & <17.4 \\
9  & J010221+015049 & 0.662 & 42300 & component\_115a & A & 5.47 & 92.8  & <6.2  & <0.067                 & <2.0  & <18.5 \\
10 & J090446+020842 & 0.794 & 55399 & component\_52a  & A & 6.76 & 191.5 & <14.0 & <0.075                 & <2.3  & <18.6 \\
    \hline
1  & J004345-022306 & 0.851    & 55394 & component\_190a & O & 5.47 & 76.1  & <8.7  & <0.11 & <3.4 & <18.8 \\
2  & J141906-020344 & 0.556    & 51446 & component\_244a & O & 4.57 & 46.0  & <2.0  & <0.044 & <1.3 & <18.4 \\
3  & J141925-014937 & 0.988    & 51446 & component\_7a   & O & 4.53 & 761.7 & <2.6  & <0.003 & <0.10 & <17.3 \\
4  & J010039-013125 & 0.604    & 42300 & component\_243a & O & 6.96 & 53.1  & <14.0 & <0.27 & <8.1 & <19.2 \\
5  & J144546-013041 & 0.536 & 52528 & component\_84a$^{\dag}$ & O & 5.17 & 167.6 & 21.0\,$\pm$\,6.4  & 0.12\,$\pm$\,0.038 & 7.5\,$\pm$\,2.1 & 19.1\,$\pm$\,0.110 \\
6  & J014316-119010 & 0.519    & 51450 & component\_15a  & O & 5.17 & 538.5 & <0.63  & <0.001 & <0.035 & <16.8 \\
7  & J090657-004243 & 0.916    & 59842 & component\_434a & O & 5.84 & 30.7  & <2.3  & <0.076 & <2.3 & <18.6 \\
8  & J141802-002314 & 0.560    & 51446 & component\_165a & O & 4.47 & 61.8  & <0.40  & <0.007 & <0.20 & <17.6 \\
9  & J143943-001854 & 0.836    & 52528 & component\_115a & O & 5.00 & 118.4 & <2.8  & <0.024 & <0.71 & <18.1 \\
10 & J141624+001639 & 0.805    & 51446 & component\_361a & O & 4.43 & 31.0  & <2.6  & <0.084 & <2.5 & <18.7 \\
11 & J143011+002220 & 0.623    & 52528 & component\_288a & O & 5.76 & 47.1  & <3.2  & <0.069 & <2.1 & <18.6 \\
12 & J085648+003920 & 0.522    & 55399 & component\_353a & O & 4.80 & 38.4  & <13.0 & <0.33 & <10.0 & <19.3 \\
13 & J090003+013541 & 0.840    & 55399 & component\_87a  & O & 5.18 & 132.2 & <7.0  & <0.053 & <1.6 & <18.5 \\
14 & J142008+014812 & 0.583    & 51446 & component\_266a & O & 4.50 & 46.0  & <2.2  & <0.049 & <1.5 & <18.4 \\
15 & J142613+020031 & 0.510    & 51446 & component\_128a & O & 5.19 & 84.4  & <2.9  & <0.035 & <1.0 & <18.3 \\
\hline  
\end{tabular}
\caption[Summary of derived \hi 21\,cm absorption properties for the 10 associated and 15 offset objects hosting bright FLASH sources]{Summary of derived \hi 21\,cm absorption properties for all 10 associated and 15 offset bright objects in this study. The table contains each galaxy name, optical redshift of the galaxy, redshift of the detection if present, the unique FLASH SBID for the cross-matched radio source observation, the component of the FLASH source cross-matched against, and the cross-match type (associated/offset). $\upsigma_{\mathrm{rms}}$ is the mean spectral line noise per channel, $S_{\mathrm{c}}$ is the peak continuum flux of the source, $\Delta S_{\mathrm{peak}}$ is the spectral line depth at either the optical redshift of the galaxy or the redshift of the absorption detection, $\tau_{\mathrm{obs}}$ is the observed optical depth, $\int \tau_{\mathrm{obs}}\mathrm{d}\nu$ is the observed velocity integrated optical depth, and the final column is column density. Upper limits are calculated assuming a single Gaussian spectral line of FWHM equal to 30\,km\,s$^{-1}$ and peak depth equal to 3$\upsigma_{\mathrm{rms}}$.}
\label{tab:bright_object_values}
\end{table}
\end{landscape}
\restoregeometry

\newgeometry{margin=1.5cm}
\begin{landscape}
\begin{table}
\renewcommand{\arraystretch}{0.8}
\centering
\begin{tabular}{c|c|c|c|c|l|c|r|c|c|c|c}
    \hline \hline
   no. & Galaxy name & \textit{z}$_{\mathrm{opt}}$ & \textit{z}$_{\mathrm{det}}$& FLASH & FLASH & $\upsigma_{\mathrm{rms}}$ & $S_{\mathrm{c}}$ & $\Delta S_{\mathrm{peak}}$ & $\tau_{\mathrm{obs}}$ & $\int \tau_{\mathrm{obs}}\mathrm{d\nu}$ & $\log_{10}\biggl(\frac{N_{\mathrm{HI}} f}{T_{\mathrm{s}}} \biggr)$ \\
   &  &  &  & SBID & component ID & (mJy\,bm$^{-1}$\,ch$^{-1}$) & (mJy) & (mJy) & & (km\,s$^{-1}$) & $\log_{10}[\mathrm{cm^{-2}\,K^{-1}}]$\\
    \hline    
1   & J142850-021443 & 0.743 & --    & 51446 & component\_305a  & 7.91 &  34.7 & <1.7  & <0.049 & <1.9 & <18.4 \\
2   & J005921-011241 & 0.504 & --    & 42300 & component\_339a  & 7.58 &  38.1 & <2.3  & <0.059 & <1.8 & <18.5 \\
3   & J142159-005244 & 0.733 & --    & 51446 & component\_261a  & 4.57 &  41.7 & <0.037  & <0.001 & <0.026 & <16.7 \\
4   & J215119-005144 & 0.598 & --    & 62514 & component\_277a  & 5.03 &  42.1 & <4.0  & <0.096 & <2.9 & <18.7 \\
5   & J214752-005114 & 0.798 & --    & 62514 & component\_77a   & 4.73 & 113.1 & <0.30  & <0.003 & <0.081 & <17.1 \\
6   &J214954-004657 & 0.864 & 0.863 & 62514 & component\_52a$^{\ddag}$  & 4.87 & 180.0 & 12.0\,$\pm$\,2.5 & 0.066\,$\pm$\,0.014 & 6.3\,$\pm$\,1.3 & 19.1\,$\pm$\,0.014 \\
7   & J090125-003703 & 0.737 & --    & 55399 & component\_23a   & 5.01 & 299.3 & <5.9  & <0.020 & <0.59 & <18.0 \\
8   & J214800-002834 & 0.515 & --    & 62514 & component\_222a  & 4.74 &  52.1 & <0.19  & <0.004 & <0.11 & <17.3 \\
9   & J143632-000726 & 0.872 & --    & 52528 & component\_243a  & 4.97 &  47.1 & <2.2  & <0.046 & <1.4 & <18.4 \\
10  & J212302+000316 & 0.462 & --    & 34578 & component\_163a  & 5.03 &  62.1 & <3.3  & <0.053 & <1.6 & <18.5 \\
11  & J091141+001101 & 0.751 & --    & 59842 & component\_5a    & 4.63 & 694.0 & <1.0  & <0.001 & <0.044 & <16.9 \\
12  & J092035+002330 & 0.924 & --    & 59842 & component\_39a   & 4.40 & 165.2 & <1.8  & <0.011 & <0.33 & <17.8 \\
13  & J145128+003426 & 0.561 & --    & 52528 & component\_292a  & 5.38 &  41.5 & <2.6  & <0.062 & <1.8 & <18.5 \\
14  & J214331+004452 & 0.457 & --    & 62514 & component\_262a  & 4.75 &  44.9 & <1.1  & <0.024 & <0.71 & <18.1 \\
15  & J010348+010621 & 0.638 & --    & 42300 & component\_90a   & 5.10 & 113.7 & <4.4  & <0.039 & <1.2 & <18.3 \\
16  & J215428+011021 & 0.598 & --    & 62514 & component\_13a   & 5.90 & 365.1 & <1.0  & <0.003 & <0.079 & <17.2 \\
17  & J090520+013103 & 0.661 & --    & 55399 & component\_215a  & 6.70 &  61.0 & <7.4  & <0.12 & <3.7 & <18.8 \\
18  & J092016+013330 & 0.689 & --    & 59842 & component\_143a  & 4.47 &  70.1 & <6.1  & <0.088 & <2.6 & <18.7 \\
19  & J142007+014809 & 0.561 & --    & 51446 & component\_266a  & 4.50 &  41.5 & <0.33  & <0.008 & <0.24 & <17.6 \\
20  & J085819+014940 & 0.625 & --    & 55399 & component\_125a  & 5.10 &  96.1 & <6.9  & <0.071 & <2.1 & <18.6 \\
21  & J085714+021018 & 0.885 & --    & 55399 & component\_299a  & 5.05 &  45.1 & <10.0 & <0.22 & <6.7 & <19.1 \\
22  & J143211+021308 & 0.528 & --    & 52528 & component\_365a  & 5.70 &  31.4 & <3.1  & <0.098 & <3.0 & <18.7 \\
23  & J143204+030543 & 0.511 & --    & 52528 & component\_16a   & 8.26 & 405.2 & <2.0  & <0.005 & <0.15 & <17.4 \\
\hline  
\end{tabular}
\caption[Summary of derived \hi 21\,cm absorption properties for the 23 spare fibre objects hosting bright FLASH sources]{Summary of derived \hi 21\,cm absorption properties for all 23 spare fibre objects hosting bright FLASH sources in this study. The table contains each galaxy name, optical redshift of the galaxy, redshift of the detection if present, the unique FLASH SBID for the cross-matched radio source observation, the component of the FLASH source cross-matched against, and the cross-match type (associated/offset). $\upsigma_{\mathrm{rms}}$ is the mean spectral line noise per channel, $S_{\mathrm{c}}$ is the peak continuum flux of the source, $\Delta S_{\mathrm{peak}}$ is the spectral line depth at either the optical redshift of the galaxy or the redshift of the absorption detection, $\tau_{\mathrm{obs}}$ is the observed optical depth, $\int \tau_{\mathrm{obs}}\mathrm{d}\nu$ is the observed velocity integrated optical depth, and the final column is column density. Upper limits are calculated assuming a single Gaussian spectral line of FWHM equal to 30\,km\,s$^{-1}$ and peak depth equal to 3$\upsigma_{\mathrm{rms}}$. $^{\dag}$ denotes where a detection has been found using the optical redshift prior.}
\label{tab:spare_radio_object_values}
\end{table}
\end{landscape}
\restoregeometry


\begin{table*}
\centering
\begin{tabular}{l|c|c|c|c|c|c}
    \hline \hline
    FLASH & $z$ & line-width & ln\,B & $\tau_{\mathrm{obs}}$ & $\int \tau_{\mathrm{obs}}\mathrm{d\nu}$ & $\log_{10}\biggl(\frac{N_{\mathrm{HI}} f}{T_{\mathrm{s}}} \biggr)$ \\
     component &  & (km\,s$^{-1}$) & & & (km\,s$^{-1}$) & $\log_{10}[\mathrm{cm^{-2}\,K^{-1}}]$\\
    \hline    
    62514\_component\_52a & 0.863 & 95 & 8.0 & 0.066\,$\pm$\,0.014 & 6.266\,$\pm$\,1.327 & 19.058\,$\pm$\,0.014 \\
    42296\_component\_52a & 0.863 & 126 & 13.5 & 0.060\,$\pm$\,0.011 & 7.518\,$\pm$\,1.300 & 19.136\,$\pm$\,0.010\\
    34555\_component\_53a & 0.863 & 138 & 22.0 & 0.069\,$\pm$\,0.010 & 9.522\,$\pm$\,1.380 & 19.240\,$\pm$\,0.010\\
    \hline
\end{tabular}
\caption[Comparison of \textsc{flashfinder} parameters assigned to absorption line of NVSS J214954-004657]{Comparison of absorption line characteristics of NVSS J214954-004657 returned by \textsc{flashfinder} for 62514\_component\_52a, 42296\_component\_52a, and 34555\_component\_53a. The first entry corresponds with the full FLASH survey observation of this source, which our work is from. The other two are the FLASH pilot survey observations of this source, which we use for comparison, and find good agreement of \textsc{flashfinder} parameters with.}
\label{tab:spare_fibre_linefinder}
\end{table*}

\begin{figure}\includegraphics[width=\linewidth]{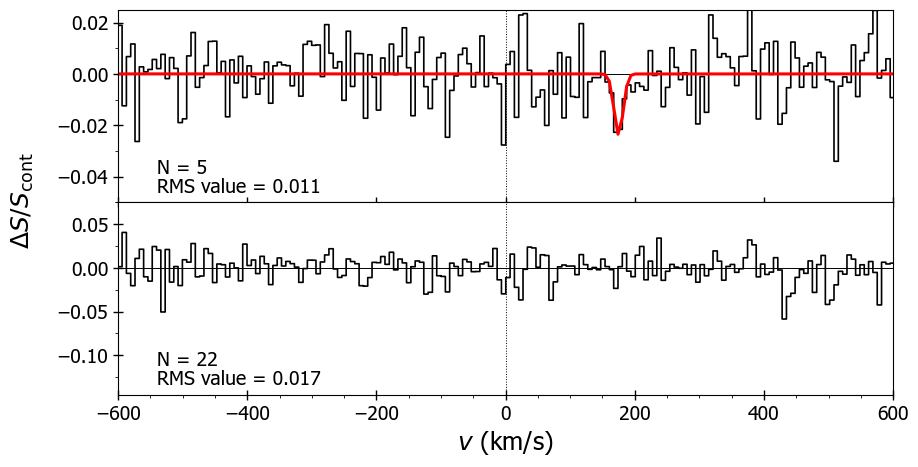}
\caption[Stacked spare fibre optical depth]{Stacked spectra produced by stacking the first 5 and then all 22 bright spare fibre FLASH spectra in terms of optical depth in order of decreasing brightness, once the spectrum from 62514\_component\_52a containing the detection is removed. The \textsc{flashfinder} detection for the $N$\,=\,5 stack is indicated in red. Values of rms opd are included for each stack. The opd rms values increase to a maximum value when all spectra have been stacked. Note that the vertical scale changes to best display the data.}
\label{fig:spare_radio_stack}
\end{figure}

\subsubsection{Spare fibre stacking search}\label{spare_fibre_stacking}

For the $N$\,=\,5 stack of Fig.\,\ref{fig:spare_radio_stack}, \textsc{flashfinder} detects a tentative \hi 21\,cm absorption signal redshifted from the central velocity of the stack, with a velocity difference of 175 km\,s$^{-1}$. The ln\,B value of this detection is 49.0, which indicates a high statistical significance. The peak observed optical depth is 0.022\,$\pm$\,0.002, the observed velocity integrated optical depth is 0.447\,$\pm$\,0.057 km\,s$^{-1}$, and the line width is 19.7\,km\,s$^{-1}$. The total background flux is 1763.69\,mJy. We do not see any further significant detections flagged for other spare fibre stacks, and as such we once again assume that the detection becomes buried as the number of spectra included in the stack increases.

The redshifted nature of this narrow detection is peculiar. With the tentative detection in the $N$\,=\,5 associated stack, the peak is centred very closely to our central redshift, which is what we would expect from a stacked detection. We again check the individual spectra of the five brightest galaxies that are stacked for weak detections which could have produced this signal, but \textsc{flashfinder} flags no detections that are close to the redshifts of the galaxies. Associated \hi 21\,cm absorption lines are known to possess broader profiles than intervening \hi 21\,cm absorption lines \citep{Curran2016b}, and if the stacked detection is astronomical in nature, we would potentially expect to see a broad stacked signal as a result. Although it is possible that a broader signal is beginning to be buried in the stacking process at $N$\,=\,5 already, we speculate that the difference in the galaxies being evaluated in the associated sample versus the spare fibre sample could be the reason for these differing absorption profiles.

The associated galaxies were selected to be UV-bright star-forming galaxies hosting radio sources, while the spare fibre galaxies are radio galaxies which we make no assumptions about the \hi content of, due to having no knowledge of whether they are actively star-forming. It is possible that associated \hi 21\,cm absorption profiles are different in galaxies known to be star-forming compared to regular radio galaxies. \citet{Gereb2014} note that the \hi 21\,cm absorption detections in their sample of radio AGN are variously blue- and redshifted between -200\,km\,s$^{-1}$ and +300\,km\,s$^{-1}$, and that the kinematics of \hi in radio galaxies must be complex. Further work will need to be done to evaluate whether the absorption profiles are different, and also the relationship between the number of spectra being stacked, the total background flux, and the resulting optical depth of detections found in stacked spectra.

As it is, we emphasise that we have no confirmation that this detection is astronomical in nature, and with the unusual redshift offset, we consider that the signal could be spurious, and thus we consider this a tentative detection only. Due to the small number of spectra in this stack, the tentative detection could result from noise fluctuations co-adding constructively in this instance. Our primary takeaway from our results for the associated, intervening, and offset stacking show that further work on \hi 21\,cm absorption stacking is necessary to isolate the regime where \hi 21\,cm absorption stacking is possible and can be a useful tool for studying neutral hydrogen at intermediate redshifts.

\begin{figure*}
\begin{multicols}{3}
    \includegraphics[width=\linewidth]{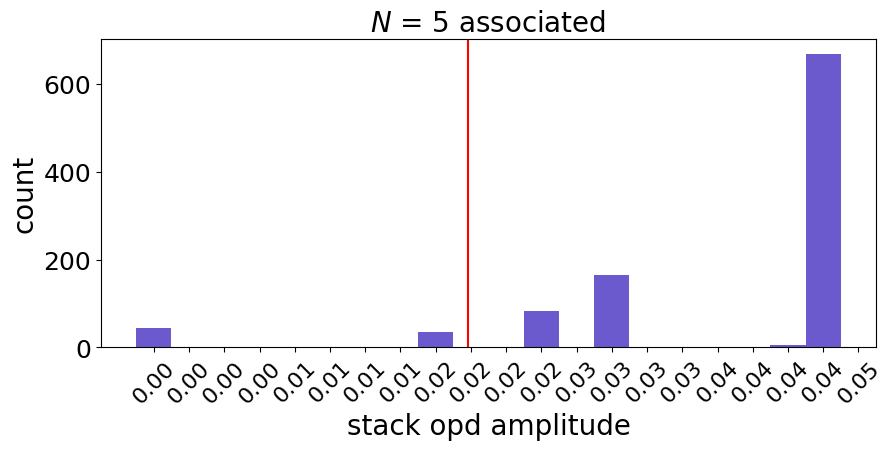}
    \par
    \includegraphics[width=\linewidth]{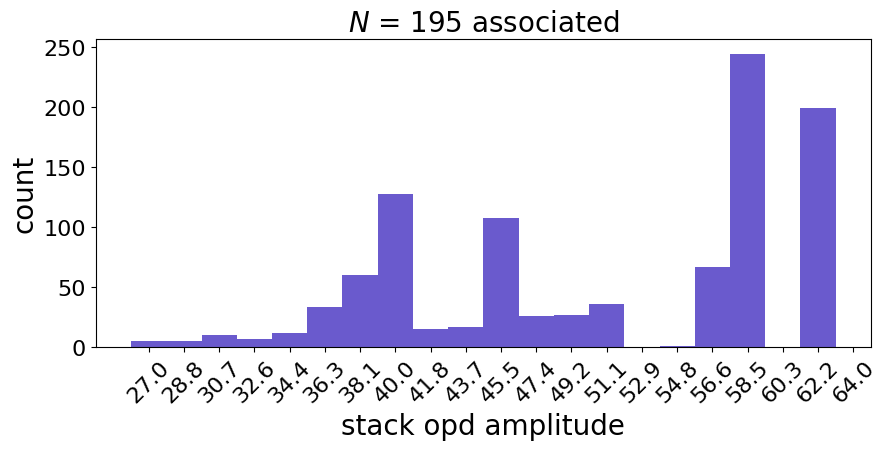}
    \par
    \includegraphics[width=\linewidth]{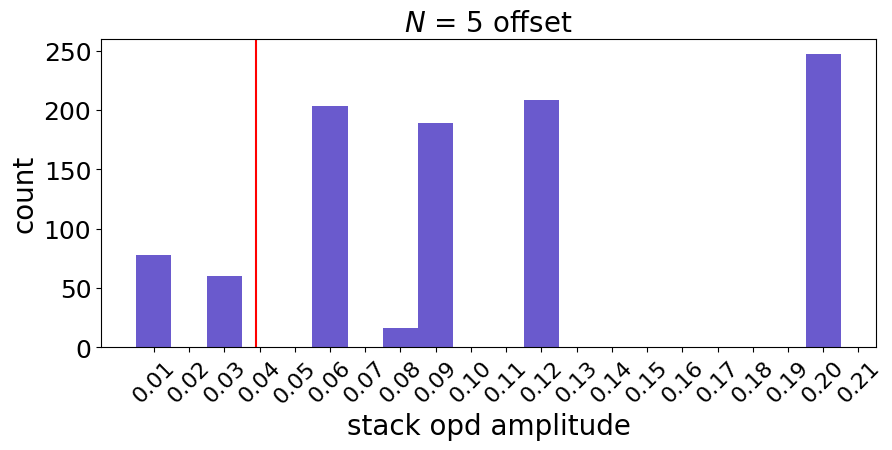}
    \par
    \includegraphics[width=\linewidth]{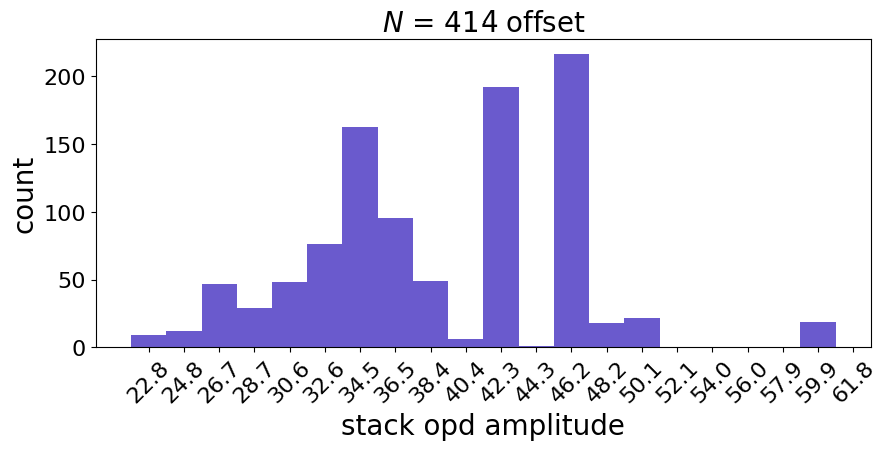}
    \par
    \includegraphics[width=\linewidth]{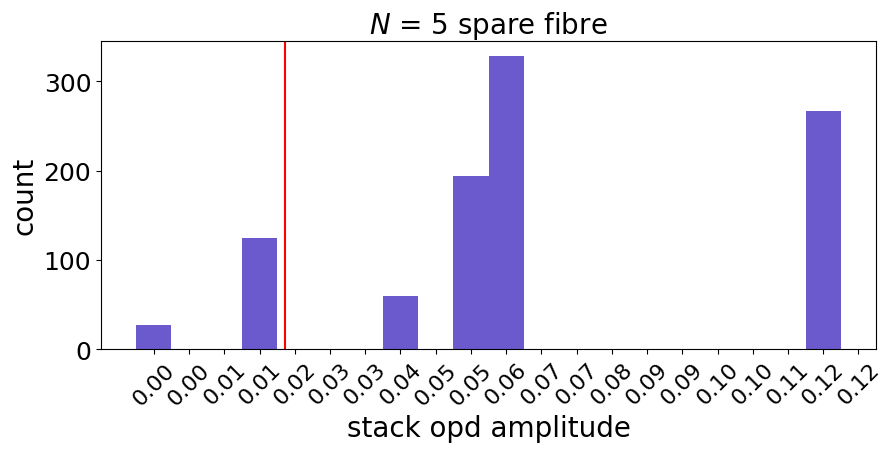}
    \par
    \includegraphics[width=\linewidth]{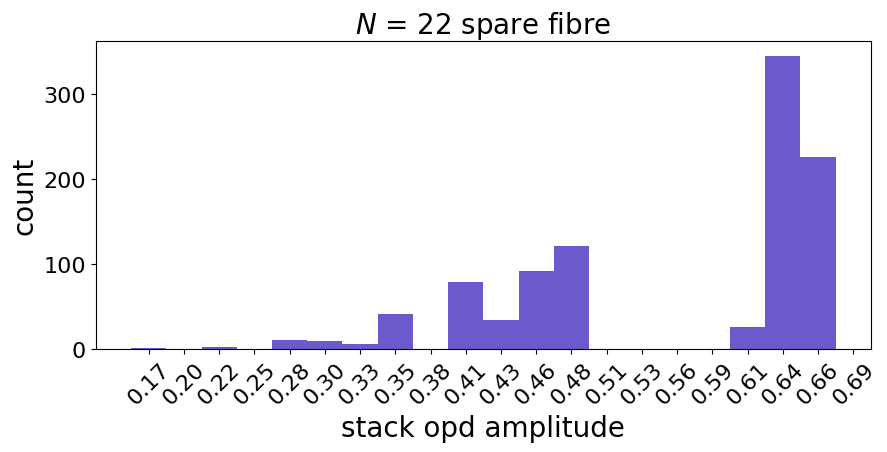}
\end{multicols}

\caption[Bootstrapping]{Bootstrapping opd results for associated, offset, and spare fibre samples, each resampled 1000 times. Bootstrapped minimum amplitude data is indicated by purple bins, while the minimum opd amplitude of the stacked spectra is indicated by a vertical red line. \textit{Left:} bootstrapping for $N$\,=\,5 and $N$\,=\,195 associated spectra. The minimum opd amplitude of the $N$\,=\,5 stacked spectrum is -0.026, while for the $N$\,=\,195 bootstrapping data, the minimum opd amplitude is -0.381, which does not overlap the bootstrapped data distribution. \textit{Centre:} bootstrapping for $N$\,=\,5 and $N$\,=\,414 offset spectra. The minimum opd amplitude of the $N$\,=\,5 stacked spectrum is -0.045, while for the $N$\,=\,414 bootstrapping data, the minimum opd amplitude is -0.161, which also does not overlap the bootstrapped data distribution. \textit{Right:} bootstrapping for $N$\,=\,5 and $N$\,=\,22 spare fibre spectra. The minimum opd amplitude of the $N$\,=\,5 stacked spectrum is -0.033, while for the $N$\,=\,22 bootstrapping data, the minimum opd amplitude is -0.051, which again does not overlap the bootstrapped data distribution.}
\label{fig:bootstrapping}
\end{figure*}

\subsubsection{Bootstrapping analysis}\label{bootstrapping}

The \textsc{linestacker} module provides capability for bootstrapping analysis of spectra stacked using it. Bootstrapping allows for the statistical evaluation of the properties of the stacked data, in particular the variance of the sample. The spectra are resampled a specific number of times, and then \textsc{linestacker} extracts information from these resampled spectra. With a large enough number of resampled spectra, conclusions can be made about the properties of the original sample. Bootstrapping does not require underlying knowledge of the distribution of the original sample. We have a relatively small number of spectra in each of our three samples, and we do not know the underlying distribution of data, so bootstrapping can allow for more robust findings to be made about a dataset.

We chose to resample the spectra $N_{\mathrm{resample}}$\,=\,1000 times, as bootstrapping is computationally expensive. The bootstrapped data was outputted into histograms by \textsc{linestacker}, and we retained the default 20 bins. We ran a modified analysis based around the minimum amplitudes of the opd in the spectra, as we are looking for absorption detections rather than emission. A set of bootstrapped data was produced for each of the samples of data used for the stacked spectra in Figs.\,\ref{fig:associated_stack},\ref{fig:offset_stack}, and \ref{fig:spare_radio_stack}, and the results can be seen in Fig.\,\ref{fig:bootstrapping}.

For all three $N$\,=\,5 bootstrapping attempts, it is difficult to draw wider conclusions about the data. The resampled data that \textsc{linestacker} returns are sparse, and the distribution is limited to certain bins. The minimum amplitudes of the stacked spectra are -0.036, -0.045, and -0.033 for the associated, offset, and spare fibre stacks respectively. These minimum opd values for the stacked spectra overlap with the distribution of the resampled data, but due to the sparse and non-normal distribution of the data, there is no clear inference that can be made about this. Importantly, this also means that the tentative stacking detections are not supported statistically.

For the $N$\,=\,195 associated, $N$\,=\,414 offset, and $N$\,=\,22 spare fibre bootstrapping attempts, the resampled data returned by \textsc{linestacker} is much better populated, although it is still difficult to draw conclusions, as the data does not follow a normal distribution. It can be seen that for all three of the full sample sizes, the distribution appears to be bimodal, with a wider collection of lower count amplitudes on the left, and a sharp peak of fewer high count amplitudes on the right. It is also important to note that for all three of these bootstrapped samples, the minimum amplitudes of the stacked spectra do not overlap with the distributions, instead lying much lower. The minimum amplitudes for the associated, offset, and spare fibre stacked spectra are -0.381, -0.161, and -0.051 respectively.

Interestingly, the resampled amplitudes returned by \textsc{linestacker} for the full sample sizes are higher than those for the $N$\,=\,5, which can especially be seen for the associated and offset stacks where the numbers of spectra in the initial samples are $\sim{10^{2}}$, similarly to the increase of rms opd values as an increasing number of spectra are co-added.

We conclude that for this investigation of bootstrapping with regards to the stacking of opd for absorption searches, there are clearly issues with interpretation of results. As it stands, the bootstrapping carried out for the tentative detections for the $N$\,=\,5 associated and spare fibre stacks, while limited, do not support the likelihood of the detections being statistically meaningful. We emphasise that the added complexity of stacking in terms of optical depth makes statistical searches via stacking and boostrapping difficult, and we suggest that important gains for 21\,cm \hi studies beyond $z$=\,0.4 can be made by developing a more robust understanding of the regimes where stacking experiments can be utilised effectively.


\section{Discussion}\label{discussion}

\subsection{Bright associated galaxies vs spare fibre galaxies}\label{bright_associated_vs_spare_radio}

As the spare fibre objects share commonalities with the bright associated WiggleZ objects, it is useful to compare them. Both sets of objects are expected to host bright FLASH sources, and we showed earlier that many of the associated WiggleZ galaxies are AGN hosts.

We find that none of the 10 bright associated WiggleZ objects contain a \hi 21\,cm absorption signal. On the other hand, one out of the 23 spare fibre galaxies contains a detection, corresponding to a detection rate of $4.3^{+1.9}_{-1.4}$ per\,cent. Although these samples are very small, applying this detection rate to the sample of bright associated objects, one would not expect to find a detection. The results from these samples are limited by their small size, and thus it would be ideal to carry out larger-scale searches in the future to determine if this detection rate would hold for both the bright associated galaxies as well as the spare fibre galaxies, and compare these results.

\subsection{Bright offset galaxies}\label{bright_offset}

Similarly to the sample of bright associated WiggleZ objects, we find that none of the 15 bright offset WiggleZ objects contain a \hi 21\,cm absorption signal. It is possible that a future search for intervening 21\,cm absorption with offset objects like the WiggleZ-type objects used in this paper could provide a way to target searches for \hi at intermediate redshifts where it has been historically more difficult to find. We have shown in \citet{Eden2025} how to produce an extended sample of WiggleZ-type objects using the Dark Energy Spectroscopic Instrument (DESI) Legacy Survey DR9 \citep{Dey2019}, Dark Energy Survey (DES) DR2 \citep{Abbott2021}, and GALEX MIS UV survey \citep{Martin2005}. With the full DESI survey DR1 \citep{Abbott2018}, a larger catalogue of WiggleZ-type galaxies covering a greater area of the sky could be produced, improving the chances of new detections.

\subsection{Comparison with previous works}\label{comparison}

\subsubsection{Direct associated searches}\label{direct_associated_searches}

\citet{Gereb2014}, conducted a search for \hi 21\,cm absorption towards 93 radio galaxies with fluxes of $S_{1.4\,\mathrm{GHz}}$ > 50 mJy, at low redshift (0.02\,<\,\textit{z}\,<\,0.23). They directly detected \hi in $\sim{29}$ per\,cent of the galaxies surveyed. In their continuation of this work, \citet{Maccagni2017} searched 248 radio galaxies with fluxes of $S_{1.4\,\mathrm{GHz}}$ > 30\,mJy, at 0.02\,<\,\textit{z}\,<\,0.25 for \hi 21\,cm absorption signals, and find a similar rate of $\sim{27\pm5}$\,per\,cent of galaxies containing direct absorption detections. On the other hand, \citet{Murthy2022} carry out a study of \hi 21\,cm absorption for 29 radio galaxies at high redshift (0.7\,<\,\textit{z}\,<\,1.0) and detect absorption towards none of their target galaxies. When they compare their \hi 21\,cm search to those done at lower redshifts, they find with 3$\upsigma$ confidence evidence that at higher redshifts, the strength of \hi 21\,cm absorptions appears to weaken. This evolution of \hi 21\,cm absorption strength with redshift is a result which has been found previously by \citet{Curran2008,Curran2019}.

At 0.4\,<\,\textit{z}\,<\,1.0 we obtain a detection rate of $4.3^{+1.9}_{-1.4}$ per\,cent in radio galaxies. From the literature, it is clear that with redshift, the detection rate of \hi 21\,cm absorption seems to decline. Our absorption detection towards NVSS J214954-004657 was assigned a ln\,B value of 8.0, which would not have been statistically significant without a redshift prior. It should be noted that the radio galaxies targeted by \citet{Murthy2022} overwhelmingly possess extended morphologies (24/29), which \citet{Curran2013b} have shown reduces \hi 21\,cm absorption detection rates, however they adjust their comparisons with low-\textit{z} surveys to take this into account.

Additionally, the radio galaxies studied in \citet{Murthy2022} have higher radio luminosities than those in lower-redshift studies such as \citet{Gereb2014,Maccagni2017}, although the sample targeted by \citet{Maccagni2017} are selected to be lower in radio luminosity than other studies at similar redshifts by around an order of magnitude, in an effort to limit the impact of radio flux on the \hi spin temperature. Thus, with our lower detection rate compared to \citet{Gereb2014,Maccagni2017}, our singular detection for 0.4\,<\,\textit{z}\,<\,1.0 does support the findings of \citet{Curran2008,Curran2019,Murthy2022} that \hi 21\,cm absorption rates seem to decline with increasing redshift, suggesting an evolution of the cool neutral medium, which could be caused by higher spin temperatures in higher redshift AGN environments.

\subsubsection{Stacked associated searches}\label{stacked_associated_searches}

While direct \hi 21\,cm absorption searches are usually successful across redshifts, \hi 21\,cm stacking experiments have faced more difficulties. \citet{Gereb2014} found and \citet{Maccagni2017} confirmed that stacking galaxies which had non-detections of \hi 21\,cm absorption at low redshifts (\textit{z} < 0.25) did not actually reveal a hidden \hi 21\,cm signal. In the only other \hi 21\,cm absorption stacking work at similar redshifts to our work, \citet{Murthy2022} also found a lack of a stacked absorption signal for 0.7\,<\,\textit{z}\,<\,1.0. In these works, various causes are discussed as potential reasons for the lack of a stacked signal, including covering factor, orientation effects, and UV photoionisation. In particular, \citet{Gereb2014} suggest that a total lack of \hi in some galaxies in their study sample has contributed to this lack of detections.

We find that \hi 21\,cm absorption stacking does potentially appear to reveal a 21\,cm absorption signal, but only under specific circumstances. As we show, when stacking spectra that do not contain \hi 21\,cm absorption detections in order of decreasing brightness, it appears that a signal can be detected in galaxies hosting radio sources (i.e. the associated and spare fibre samples), but only when small numbers of the spectra from the brightest sources from these samples are stacked. In both the associated and spare fibre stacking cases, we find that when $N$\,=\,5 - i.e. when the five brightest galaxy spectra are stacked - \textsc{flashfinder} detects an absorption signal with a high statistical significance.

We do not believe this is a function of $N$, but a function of background radio source brightness. That is to say, that when too many weak background sources are co-added into the stacked spectra, the optical depth limit of the stack is driven up as the median source brightness is brought down. Put another way, the sample does not reach uniform optical depth limits, so adding in spectra which do not meet this optical depth limit only introduces more noise into the stack, and does not help to reveal a stacked absorption signal.

\citet{Gereb2014,Maccagni2017,Murthy2022} measure the 3$\upsigma$ optical depth limits of their stacked samples of non-detections, and find values of $\tau$\,<\,0.002, $\tau$\,=\,0.0015, and $\tau$\,$\approx{0.0017}$. It is readily apparent that these three separate experiments reach similar optical depth limits. \citet{Gereb2014} also produce a stack of galaxies containing \hi 21\,cm absorption detections, resulting in $\tau$\,=\,0.02. In comparison, for our tentative associated and spare fibre detections, we get optical depths of $\tau$\,=\,0.006\,$\pm$\,0.001 and $\tau$\,=\,0.022\,$\pm$\,0.002 respectively, much closer to the optical depth value for stacked \hi 21\,cm absorption detections in the literature than stacked non-detections.

Further work will need to be done to determine the most optimal way to deal with the effect of optical depth on \hi 21\,cm absorption stacking. In the future, it may be possible to improve optical depth limits by stacking more spectra of bright sources with known spectroscopic redshifts, or by using spectra from a larger and more sensitive telescope such as the SKA, to lower the noise on individual spectra that are to be co-added into the stack. Alternatively, a survey of objects using the SKA designed to reach uniform optical depth limits rather than uniform noise could help to remove the issue with co-adding in more objects into the stacks.


\subsubsection{Direct intervening searches}\label{direct_intervening_searches}

Searches for intervening \hi 21\,cm absorption at intermediate redshifts have generally relied on targeting galaxies with bright radio sources in their background that exhibit other types of absorption systems, such as Mg\,\textsc{ii} \citep{Curran2007,Gupta2012,Dutta2020}, which is used as a tracer for neutral gas \citep{Rao2000}. Surveys like FLASH are now allowing for blind searches for intervening \hi 21\,cm absorption given their sensitivity and the large area covered \citep{Sadler2020}. In this work, we target star-forming galaxies using the star-forming activity as a proxy for neutral gas. We seek to understand whether this approach to searching for intervening absorption could be useful for increasing the known number of intervening absorbers at higher redshifts.

In our sample of 15 offset galaxies cross-matched with bright radio sources, we do not detect \hi 21\,cm absorption. This detection rate is low considering previous works such as \citet{Gupta2010,Zwaan2015} determine a detection rate of 40--50\,per\,cent for galaxy-quasar pairs. However, \citet{Reeves2015,Reeves2016} find a decreased detection rate of $\sim{6}$\,per\,cent instead towards gas rich galaxies at $z$\,<\,0.04, although their galaxy sample faces the issue of \hi 21\,cm emission possibly diluting any absorption.

This is interesting considering the different redshift regimes we are probing, and the fact that we also do not constrict the background radio source type to quasars. However, we also do not know how many of the radio sources are truly in the background of our samples, or what the covering factors in our work are. In the process of this work, we have assumed that all radio sources are likely to be in the background of our galaxies with high covering factors, but as our intervening galaxy sample is at intermediate redshift, this is likely not the case.

\citet{Reeves2016} show that their low \hi 21\,cm absorption detection rate is greatly affected by sightlines of background radio sources not actually intersecting the \hi discs of their target galaxy, with 14/23 (60.1\,per\,cent) of their sample exhibiting this. In some cases, this lack of intersection is affected by disc orientation, but primarily they explain that for their sample of radio sources, impact parameters between the \hi disc and radio source are simply too large for a detection to be made. They estimate that for sources with brightness $\geq{50}$\,mJy and within an impact parameter of 20\,kpc, the expected detection rate would not exceed 10\,--\,20\,per\,cent. Importantly, they note that the selection criteria for their sample matches closely with that of the FLASH survey, and as such they expect that their results would be representative of the expected detection rate using FLASH sources. As we have stated, our associated sample has an impact parameter of $<$\,30\,kpc, while the offset sample impact parameters are between 30\,--\,120\,kpc. Considering this, the lack of detections in our offset sample is likely primarily driven by impact parameters which are too large to physically allow detections.

Beyond this, due to the small sample sizes in this work, and that of \citet{Reeves2016}, further research with larger sample sizes would be needed before definitive conclusions on whether the detection rate of intervening \hi 21\,cm absorption really does remain consistent across redshift can be made.

\subsubsection{Stacked intervening searches}\label{stacked_intervening_searches}

Stacking of intervening \hi 21\,cm absorption spectra has been attempted, however this is at low redshift (\textit{z}\,$\le$\,0.35), with 11 systems known to contain intervening \hi \citep{Hu2025}. In this work, we stack non-detections of intervening \hi 21\,cm absorption for the first time. Unlike our results for the associated and spare fibre samples, we do not see any indication of detected absorption in any of the stacked spectra. This seems to agree generally with the findings of \citet{Hu2025}, where they show that their stacked associated absorption spectrum has a greater column density than their stacked intervening absorption spectrum, suggesting that on average the column density of associated absorption is greater than intervening absorption. 

We consider that issues with covering factors are likely compounded in stacking intervening absorbers. In Section\,\ref{stacked_associated_searches}, we discuss that the background source brightness and optical depth seem to affect both the associated and spare fibre stacks. As weaker background radio sources are co-added into the stack, the optical depth is driven up. In the intervening absorption setup, bright background sources may not be fully covered by the foreground galaxy, in a sense reducing the brightness of the radio source being co-added into the stack. This is difficult to remedy, as usually covering factors are assumed to be equal to 1 for absorption searches. This an acceptable assumption for direct searches, but for stacking searches such as this the issue will compound.

Additionally, as with the searches of individual intervening galaxy spectra, we assume that the radio sources lie in the background of the intervening galaxies, but this is not always the case. Photometric redshift distributions for FLASH sources have been estimated, finding $\sim{13}$\,per\,cent lie in the foreground at $z$\,<\,0.42, $\sim{35}$\,per\,cent within the detectability range of FLASH ('in-band'), and $\sim{52}$\,per\,cent in the background at $z$\,>\,1 \citep{Roster2026}. While more than half of FLASH sources are at redshifts high enough to be in the background of any of the galaxies we are targeting, we do not know which of the sources used in this work are background sources.

For intervening galaxy stacking in \hi 21\,cm absorption to be practical, improved ability to ensure radio sources lie in the background of intervening galaxies with more accurate photometric redshifts for radio sources would be an important step, but would not help mitigate the difficulties with covering factors.


\subsection{The critical UV luminosity}\label{critical_luminosity}

\citet{Curran2012b} discuss the comparative lack of \hi 21\,cm absorption detections for \textit{z}\,$\geq${1.0} compared to \textit{z}\,$\leq${1.0}. Crucially, although at higher redshifts one would expect to be looking at objects in the Universe at an earlier stage in their evolution - and thus that these objects would contain a much greater quantity of unconsumed \hi \citep{Peroux2001} - in practise absorbers are found at higher rates below \textit{z} = 1.0 than above \citep{Curran2013a,Curran2018}. This is especially interesting considering the findings of \citet{Chowdhury2022} where they carry out \hi 21\,cm emission stacking for 11,419 star-forming galaxies at 0.74\,<\,$z$\,<\,1.45, and find that the average \hi mass in these galaxies is $\sim{3-4}$ times higher than the average \hi mass of galaxies at $z$\,=\,0.

There has been evidence of a potential critical UV luminosity affecting detections of \hi via the 21\,cm line for some time, as in their search for \hi 21\,cm absorption, \citet{Curran2008} found no associated detections in galaxies hosting sources of UV where in the rest frame their UV luminosity at $\uplambda$ = 1216\,$\Angstrom$ is $L_{1216} \geq{10^{23} \mathrm{W\,Hz^{-1}}}$. Similarly, they found that in objects searched for intervening absorption with bright background sources, if the background source UV luminosity at the rest frame of the intervening object is also $L_{1216} \geq{10^{23} \mathrm{W\,Hz^{-1}}}$, then a detection of intervening \hi 21\,cm absorption would not be made.

\citet{Curran2012b} derive not only an associated critical photoionisation rate of $Q_{\mathrm{HI}} \sim{3 \times 10^{56}}\,\mathrm{s}^{-1}$, but also a galactic scalelength over which this ionisation will be effective, finding that a large spiral galaxy would be completely ionised. Further, in \citet{Curran2017}, they find that there is significant ($\sim{7}\upsigma$) evidence for this critical UV luminosity, above which all \hi is ionised, and 21\,cm absorption is unable to be detected.

If either the associated and spare fibre host galaxies, or the bright radio sources in the background of our offset galaxies possess this critical UV luminosity, then the lack of detections in our subsamples are explained directly. On the other hand, if these samples do not exceed this critical UV luminosity, then some other combination of effects must be responsible, such as orientation effects, low covering factors, survey sensitivity, or in some cases a genuine lack of \hi gas.

\subsubsection{UV luminosity results}\label{uv_luminosity_results}

We calculate the ionising photon rate for the associated, offset, and spare fibre galaxy samples. The necessary data is available for 103/209 (49.3 per\,cent) of the associated objects and 227/454 (50.0 per\,cent) of the offset objects. The spare fibre galaxies do not contain as complete SDSS photometry due to not being targeted by WiggleZ selection criteria, and thus only 59/398 (14.8 per\,cent) of the total number of these galaxies contains the required photometry to calculate values.

\begin{figure}
    \centering
    \includegraphics[width=\linewidth]{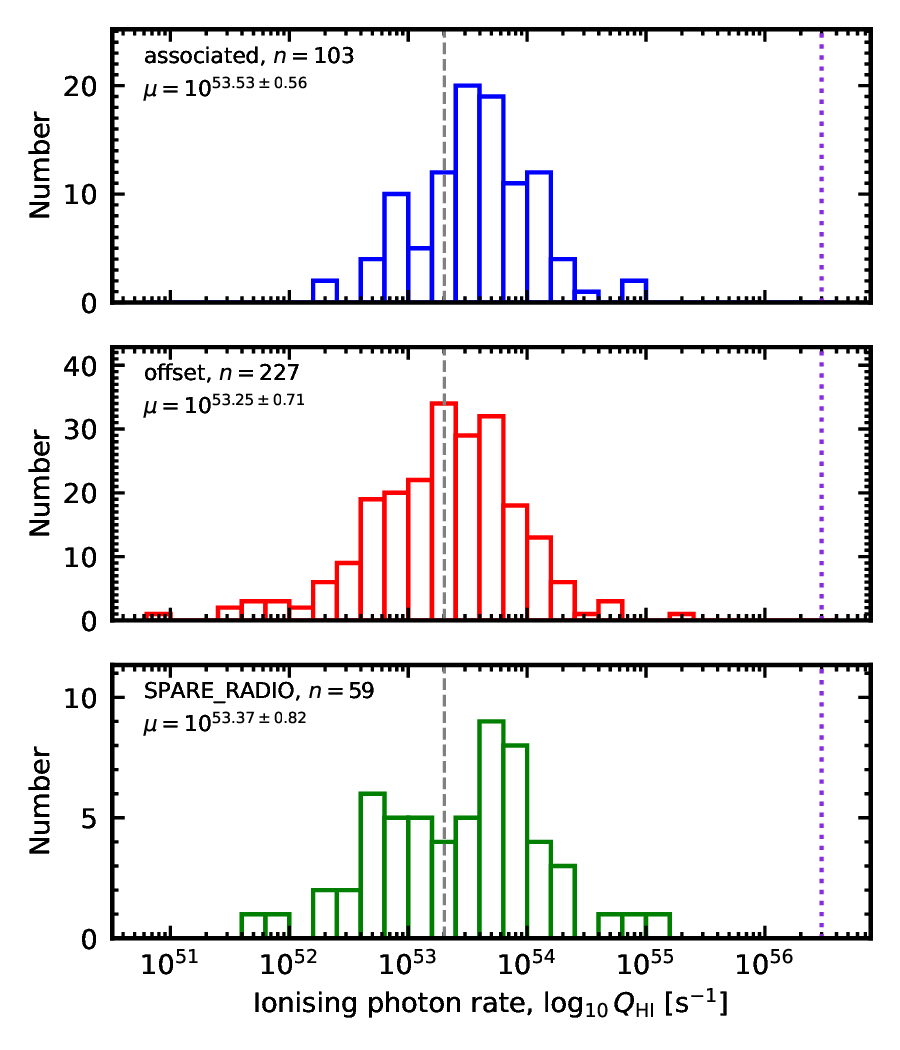}
    \caption[Distribution of ionising photon rates for associated, offset, and spare fibre galaxies]{Comparison of the distribution of ionising photon rates for associated, offset, and spare fibre galaxies. Sample size and mean rate with uncertainty for each sample is included. The median UV luminosity for each sample is indicated with the vertical grey dashed line, and the critical UV luminosity with the vertical purple dotted line.}
    \label{fig:ionising_photon_rate_distribution}
\end{figure}

We find for all three samples that the maximum ionising photon rate is below that of what would be expected to ionise \hi gas within the galaxies significantly enough to be the main cause of non-detections. The maximum rate for the associated sample is $Q_{\mathrm{HI}} = {7.5 \times 10^{54}}\,\mathrm{s}^{-1}$, for the offset sample is $Q_{\mathrm{HI}} = {1.7 \times 10^{55}}\,\mathrm{s}^{-1}$, and for the spare fibre sample is $Q_{\mathrm{HI}} = {1 \times 10^{55}}\,\mathrm{s}^{-1}$ (see Fig.\,\ref{fig:ionising_photon_rate_distribution} for full distribution). With rates of $Q_{\mathrm{HI}} \lesssim{1 \times 10^{55}}\,\mathrm{s}^{-1}$, these are all within the range where \hi is typically detected \citep{Curran2019}. Further, we find that the associated sample generally has double the ionising photon rate of the offset sample - this is expected, as the associated galaxies are more likely to be AGN hosts than the star-forming offset galaxies.

We assume that these values are representative of the samples as a whole. In this case, all of the sources possess ionising photon rates lower than the highest at which \hi 21\,cm absorption has been found \citep{Curran2024}. The ionisation of the \hi gas in these galaxies cannot be the main limiting factor of the lack of detections in our samples. Instead, we look towards a range of other issues.

\subsection{Limiting factors}\label{limiting_factors}

If the degree of ionisation of the \hi in the target galaxies is not the primary issue in making detections of \hix, then other factors must be responsible. Issues we will consider are covering factor, geometric effects, depletion of \hi gas, and the impact of survey sensitivity.

The covering factor ($f_{\mathrm{c}}$) of a radio source details the fraction of the radio source covered by \hi gas. If a radio source isn't completely covered, then the strength of a \hi 21\,cm absorption detection may be reduced, or become undetectable. Whether a radio source is compact or extended can further affect this, with compact sources being more likely to have higher value of $f_{\mathrm{c}}$ than extended sources. This is discussed in \citet{Gereb2014}, where they explain that while compact sources are more likely to have a higher covering factor, this doesn't mean that the values have to be large \citep[for example $f_{\mathrm{c}}\sim{0.2}$ in B2352+495;][]{Araya2010}. This also holds for extended sources, where the covering factor can be high \citep[for example $f_{\mathrm{c}}\sim{1}$ in 3C\,293;][]{Beswick2004}. Alternatively, the \hi in galaxies can be distributed in a non-uniform manner, with clouds of higher densities embedded in a more diffuse medium \citep{Reeves2015}, which can reduce the covering factor.

Given we know that $\sim{10}$ per\,cent of our cross-matched sample is matched against an extended source \citep{Eden2025} and the high redshifts of the sample, it is reasonable to assume that the covering factor of the objects we are investigating for \hi 21\,cm absorption signals is partially responsible for the lack of detections we have found \citep{Curran2018}.

Orientation effects are also likely to have affected our detections. In the local group at z $\sim{0}$, the orientation of \hi discs in relation to the observer can be studied in great detail \citep[for recent detailed studies, see][]{Kurapati2020,Eibensteiner2023}, but at redshifts above this the inherent faintness of 21\,cm emission means that \hi disc structure cannot be observed. When searching for intervening absorption at lower redshifts (\textit{z} < 0.04), successful detections tended to occur in galaxies oriented face-on or close to face-on to the observer \citep{Curran2016a}.

When considering the impact of \hi gas depletion on the galaxies we are probing, it is not likely that this will have affected our results. Our galaxies are selected to be actively star-forming, built from WiggleZ galaxies and an extended sample that in \citet{Eden2025} we showed closely match the characteristics of the WiggleZ galaxies. Because of the close link between \hi abundance and star formation, these galaxies are expected to contain \hix, and as such depletion of \hi is not anticipated to have an impact on our results.

\subsubsection{Geometric effects}\label{geometric effects}

Until now, we have assumed that for intervening \hi 21\,cm searches, even though we are working with galaxies at 0.4\,<\,\textit{z}\,<\,1.0, all radio sources we have cross-matched against would lie in the background of the star-forming galaxies, at higher redshifts. Unfortunately, we are not able to investigate this as we do not have spectroscopic redshift values for the radio sources, although we know that $\sim{52}$\,per\,cent of FLASH sources are at $z$\,>\,1, and $\sim{35}$\,per\,cent at  0.42\,<$z$\,<\,1.0. In a simulation of the radio sky, \citet{Wilman2008} find similar results, with $\sim{65}$\,per\,cent of radio sources exising at \textit{z}\,>\,1, and a further $\sim{20}$\,per\,cent at 0.5\,<\,\textit{z}\,<\,1.0. Being conservative, we can assume that roughly only half of the radio sources that we have cross-matched against exist at a higher redshift than the target galaxies. In the half of cases where the radio source is at a lower redshift than the target galaxy, no \hi 21\,cm absorption would be possible to detect.

Additionally, the structure and angular size of the continuum source is very important in the detection of \hi 21\,cm absorption. If the angular size of a continuum source is much larger than that of a \hi cloud, then $f_{\mathrm{c}}$ is small. \hi clouds are typically 100--200\,pc in size \citep{Braun2012}, which at the typical WiggleZ-type galaxy redshift \citep[\textit{z} $\sim{0.6}$;][]{Eden2025} would equate to <\,0.1\,arcsec. Only $\sim{30}$\,per\,cent of galaxies are expected to have a compact component that small \citep{Chhetri2018}. On the assumption that most radio sources will be larger than the \hi clouds, then $f_{\mathrm{c}}$ < 1 for the majority of galaxy-source pairs that are well-aligned at these redshifts.

As mentioned previously, line of sight is a critical factor in the detection of \hi 21\,cm absorption at low redshifts. At low redshifts (\textit{z}\,<\,0.04), \citet{Reeves2016} determined that even in galaxies known to be \hix-rich, sightlines between background radio sources only intersected the \hi disc of the galaxy $\sim{30}$\,per\,cent of the time.

Given all of these effects, it is reasonable to conclude that the geometric setup during \hi 21\,cm absorption searches at intermediate redshifts is highly influential on whether \hi can be detected. Larger sample sizes are needed in order to increase the number of 21\,cm absorption detections that can be made.

\subsubsection{Survey sensitivity}\label{survey_sensitivity}

\begin{figure}
    \centering
\includegraphics[width=\linewidth]{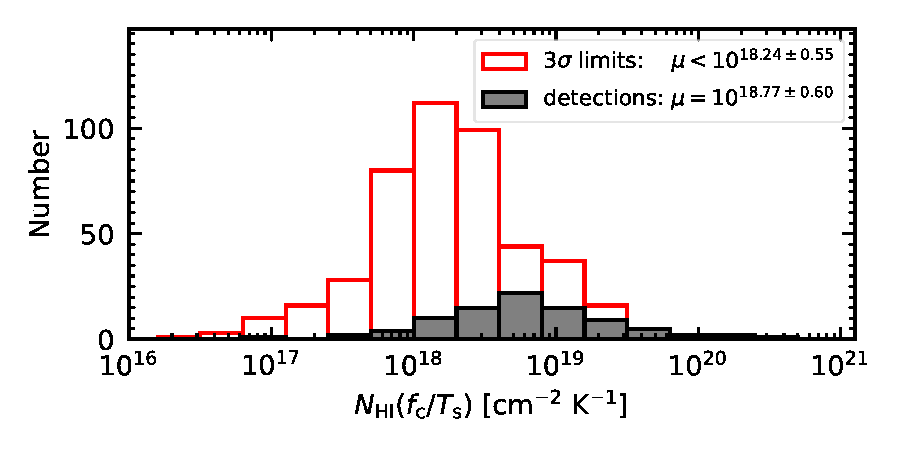}
    \caption[Distribution of line strength limits]{Distribution of line strength limits and comparison against detections from the literature \citep[excluding][]{Su2022}. The red bars are $3\upsigma$ limits of non-detections from this work, and the filled black bars are from detections in the literature.}\label{fig:line_strength_comparison}
\end{figure}

Another limiting factor is sensitivity of observations made with FLASH. Each FLASH field has an integration time of around two hours \citep{Yoon2025} compared to deeper ASKAP surveys such as the Evolutionary Map of the Universe \citep[EMU;][]{Norris2021}, for which fields have an integration time of around ten hours each. It is possible that we are reaching the limits of the FLASH survey in this search given its two hour integration times, and that for our sample size of cross-matched objects we have reached the maximum number of detections that could be expected to be found for this search. We determine sample limits for the direct detection rates of our samples and compare them to those found in other comparable surveys, namely \citet{Vermeulen2003,Gereb2015,Maccagni2017,Murthy2021,Su2022,Aditya2024}.

The channel width of FLASH spectra is $\upDelta\nu$ = 18.5\,kHz, and for the FLASH frequency range (711.5\,MHz -- 999.5\,MHz) this results in a spectral resolution of $\upDelta v$ = 5.5 -- 7.8\,km\,s$^{-1}$. In order to compare these values with literature values, these are resampled to $\upDelta v$ = 20\,km\,s$^{-1}$ which is used as the full-width half-maximum (FWHM) for calculating the integrated optical depth limits per channel \citep{Curran2012a}. This is only possible for objects where $\upsigma_{\mathrm{rms}} \lesssim{S_{\mathrm{c}}/3}$. As the typical FLASH rms spectral noise is $\sim{5}$\,mJy\,bm$^{-1}$, in many cases $3\upsigma_{\mathrm{rms}} > S_{\mathrm{c}}$ meaning normalised line strength limits could be determined for 219 of the objects (from 29 associated, 70 offset, and 120 spare fibre samples combined). These $3\upsigma$ limits are plotted against literature values in Fig.\,\ref{fig:line_strength_comparison} excluding \citet{Su2022}, due to their selection of very faint continuum sources ($S_{\mathrm{c}}$\,$\lesssim$\,10\,mJy) which may not have been searched significantly deeply to detect \hi absorption \citep[see fig.\,2 of][]{Curran2024}. The mean $3\upsigma$ limit for \citet{Su2022} is $\mu\,<\,19.64\,\pm\,0.56\,\mathrm{cm^{-2}\,K^{-1}}$ \citep{Curran2024}, and would skew the comparison of the line strength limit for this work versus the literature. Considering this, it is clear that the normalised line strength limits calculated for the objects considered in this work are weaker than those of detections in the literature. However, the depth of the search is possibly deep enough to be compared to these detections \citep{Curran2024}.

\begin{figure}
    \centering
    \includegraphics[width=\linewidth]{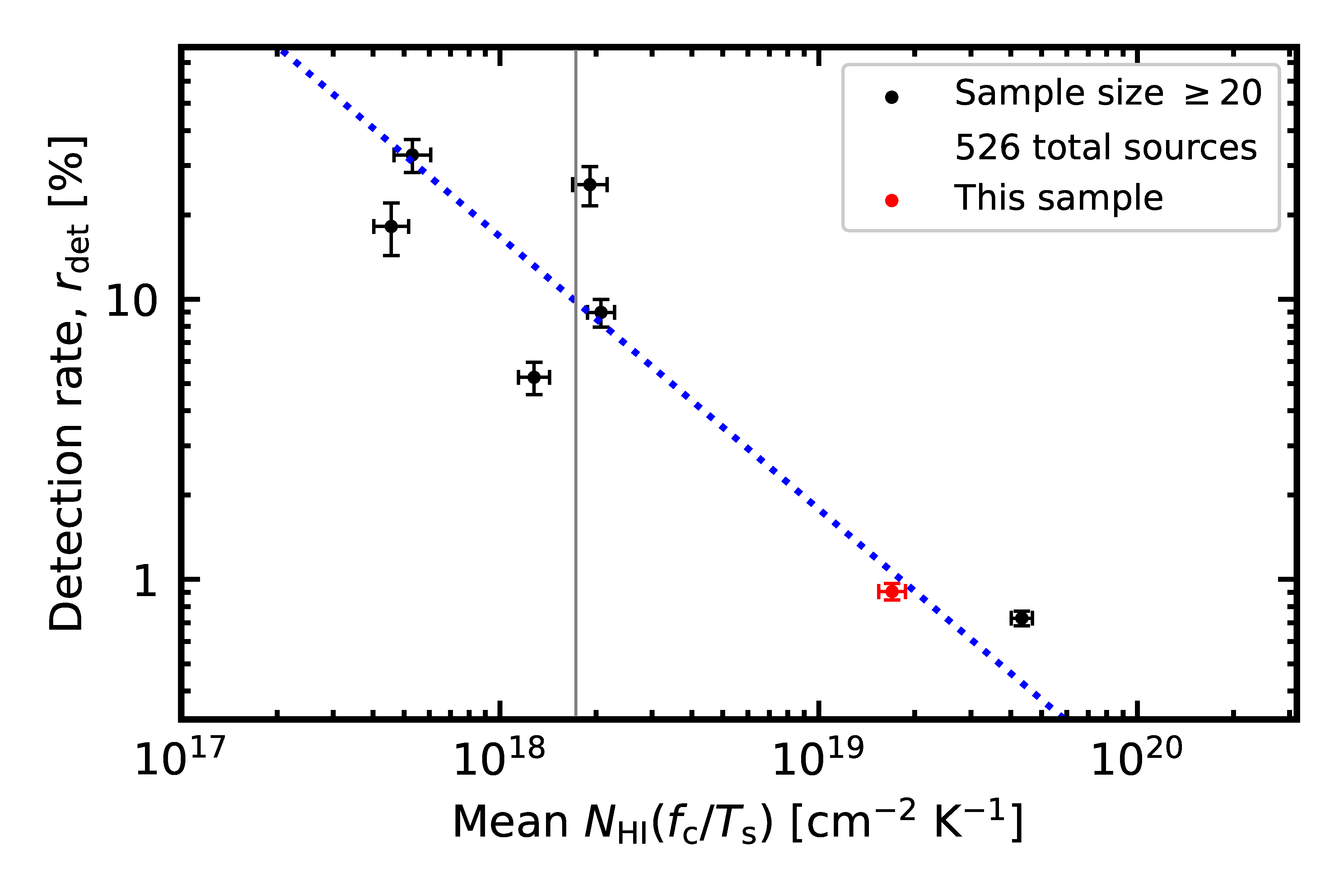}
    \caption[Detection rate versus mean \hi 21\,cm absorption strength]{Detection rate versus mean \hi 21\,cm absorption strength for surveys that contain at least 20 sources at \textit{z} $\le{1}$ \citep[black points;][]{Vermeulen2003,Gereb2015,Maccagni2017,Murthy2021,Su2022,Aditya2024}. The red point shows this survey, and the vertical grey line shows the mean $3\upsigma$ sensitivity of this survey. The bars show the standard error, and the dotted blue line shows the fit to the literature values (black error bars).}
\label{fig:detection_rates}
\end{figure}

To quantify the expected detection rate, we split the literature into the respective surveys and determine the detection rate of each. The surveys are restricted to redshifts of \textit{z} $\le{1}$ to minimise biases due to photoionisation, and must include at least 20 objects. This is in order to provide a good compromise between the number of binned data points and their quality. We present the detection rate against mean line strength in Fig.\,\ref{fig:detection_rates}. 

From this, we see an apparent anti-correlation between the detection rate and the mean line strength, which appears to be follow a power law. This is largely driven by the end point \citep[at $\langle\log_{10}N_{\mathrm{HI}}(f_{\mathrm{c}}/T_{\mathrm{s}})\rangle$ = 19.64,][]{Su2022}, which comprises 275 sources searched in total. Fitting the literature values via orthogonal distance regression to weigh by the uncertainties, we find $\log_{10}r_{\mathrm{det}} \approx{-\log_{10}\langle N_{\mathrm{HI}}(f_{\mathrm{c}}/T_{\mathrm{s}})\rangle + 19}$, where $r_{\mathrm{det}}$ is the detection rate in per\,cent. This gives an expected $r_{\mathrm{det}}$ = 0.9\,$\pm$\,4\,per\,cent for $\langle \log_{10}N_{\mathrm{HI}}(f_{\mathrm{c}}/T_{\mathrm{s}})\rangle$ = 19.23 (shown by the red point in Fig.\,\ref{fig:detection_rates}). For $N$\,=\,219 objects, this is consistent with the one associated detection. That is, presuming that the six data points (comprising 526 sources) provide a reliable measure of the $r_{\mathrm{det}}-\langle N_{\mathrm{HI}}\rangle$ relation, our associated spare fibre detection is what would be expected to be found from a search like ours with FLASH's sensitivity and the object sample sizes.

With this search we are nearing the sensitivity limits of FLASH. It is possible that there are other \hi absorption signals in our samples that are unable to be detected due to being buried in noise in the spectra. Other more sensitive telescopes, such as MeerKAT \citep{Jonas2009}, or the upcoming SKA \citep{Dewdney2009}, with their higher sensitivities than ASKAP, could be able to uncover these signals.


\section{Conclusion}\label{conclusion}

We have investigated our associated, offset, and spare fibre galaxy samples for direct \hi absorption as well as conducting a 21\,cm absorption stacking search, with ASKAP-FLASH. These galaxies lie in 0.4\,<\,\textit{z}\,<\,1.0. The associated and offset galaxies are UV-bright with ongoing star-formation, meaning that we expect them to have large \hi reservoirs. The spare fibre galaxies are radio galaxies in the same redshift range, but are not UV-selected, which we use as a control sample for the associated galaxies. We present a summary of our findings below:

\begin{enumerate}[i.]

\item
\textsc{flashfinder} identifies candidate absorption-like features in the $N$\,=\,5 associated and spare fibre stacks, but bootstrap resampling does not support these as robust stacked detections. We therefore treat them as candidate features and use them primarily to illustrate the sensitivity and optical-depth limitations of stacking at these redshifts. Having stacked in order of descending brightness, we hypothesise that the effect of apparent stacked absorption signals being flagged at $N$\,=\,5 but not at higher numbers of total stacked spectra may be due to optical depth in the stacked spectra increasing as weaker background sources are co-added into the stack, leading to the signal becoming undetectable for higher values of $N$. This may be the reason for the lack of \hi 21\,cm absorption detections via stacking in the literature.

\item
We find a new associated \hi 21\,cm absorption detection towards NVSS J214954-004657 in the spare fibre sample. This detection was found using the galaxy's spectroscopic redshift (\textit{z}$_{\mathrm{opt}}$ = 0.863) as a prior, and has a velocity difference of -160\,km\,s$^{-1}$. The Bayesian likelihood of this detection determined by \textsc{flashfinder} is ln\,B $\sim{8.0}$, which signifies that the detection is likely to be real when found at a redshift prior. The detection has a peak observed optical depth of 0.066\,$\pm$\,0.014, and an integrated optical depth of 6.266\,$\pm$\,1.327\,km\,s$^{-1}$, implying that the detection has a significance of 4.7$\upsigma$. This feature was confirmed in the previous FLASH observation of the component.

\item We do not determine a significant difference in the \hi 21\,cm absorption detection rate for star-forming galaxies compared to the general population of radio galaxies at intermediate redshifts.

\item
In addition, we have carried out statistical analyses of the detection rates of our search. We find that the expected number of detections from our sample size when compared to similar \hi 21\,cm absorption searches in the literature is consistent with our singular detection. 

\end{enumerate}

\noindent
As we approach the SKA era, searches for \hi at 0.4\,<\,\textit{z}\,<\,1.0 will soon be able to probe for detections at greater sensitivities than ever before. This will allow for detections of \hi 21\,cm absorption at much lower line strengths than has previously been possible, opening up the potential for tracing \hi at these intermediate redshifts where searches have been limited to galaxies associated with or in the foreground of the brightest radio sources.

In this series, we have produced a large sample of intermediate redshift star-forming galaxies, and investigated their \hi content. As FLASH is designed for detecting \hi 21\,cm absorption, and due to the star-forming nature of the galaxies we have searched, we expected to detect a higher number of associated and intervening \hi 21\,cm signals, as well as evidence of \hi 21\,cm stacking searches in complete stacks of non-bright spectra. Instead, we have learnt that searches of these star-forming galaxies for direct detections pushes the limits of what FLASH is capable of as a survey, as well as that absorption stacking is potentially possible, but seems to be limited by a dependence on optical depth and background radio source strength.

Our results demonstrate that the improvements in capability that SKA will bring (sensitivity, higher angular resolution), are necessary to continue to develop our understanding of \hi at these redshifts. We suggest that further work investigating \hi 21\,cm absorption stacking with SKA would be of benefit in understanding the limits placed by optical depth and radio source brightness. This will allow a greater understanding of \hi content at intermediate redshifts by including statistical approaches to 21\,cm absorption searches at these redshifts for the first time. We also suggest that with larger sample sizes and greater sensitivity of SKA, it may be possible to search for stacked 21\,cm absorption by redshift bin, which could allow for tracing of the evolution of \hi statistically during this crucial period of cosmic time.


\section*{Acknowledgements}

SLE thanks STFC for support through a STFC doctoral scholarship. 

This scientific work uses data obtained from Inyarrimanha Ilgari Bundara / the Murchison Radio-astronomy Observatory. We acknowledge the Wajarri Yamaji People as the Traditional Owners and native title holders of the Observatory site. CSIRO’s ASKAP radio telescope is part of the Australia Telescope National Facility (https://ror.org/05qajvd42). Operation of ASKAP is funded by the Australian Government with support from the National Collaborative Research Infrastructure Strategy. ASKAP uses the resources of the Pawsey Supercomputing Research Centre. Establishment of ASKAP, Inyarrimanha Ilgari Bundara, the CSIRO Murchison Radio-astronomy Observatory and the Pawsey Supercomputing Research Centre are initiatives of the Australian Government, with support from the Government of Western Australia and the Science and Industry Endowment Fund.

This work was supported by the Australian SKA Regional Centre (AusSRC), Australia’s portion of the international SKA Regional Centre Network (SRCNet), funded by the Australian Government through the Department of Industry, Science, and Resources (DISR; grant SKARC000001). AusSRC is an equal collaboration between CSIRO – Australia’s national science agency, Curtin University, the Pawsey Supercomputing Research Centre, and the University of Western Australia.

Parts of this research were supported by the Australian Research Council Centre of Excellence for All Sky Astrophysics in 3 Dimensions (ASTRO 3D), through project number CE170100013. 

JH acknowledges support from the UK SKA Regional Centre (UKSRC). The UKSRC is a collaboration between the University of Cambridge, University of Edinburgh, Durham University, University of Hertfordshire, University of Manchester, University College London, and the UKRI Science and Technology Facilities Council (STFC) Scientific Computing at RAL. The UKSRC is supported by funding from the UKRI STFC.

Part of the research activities described in this paper were carried out with contribution of the Next Generation EU funds within the National Recovery and Resilience Plan (PNRR), Mission 4 - Education and Research, Component 2 - From Research to Business (M4C2), Investment Line 3.1 - Strengthening and creation of Research Infrastructures, Project IR0000034 – ``STILES - Strengthening the Italian Leadership in ELT and SKA''

HY is supported by the National Research Foundation of Korea (NRF) grant funded by the Korea government (MSIT, RS-2025-00516062), and by funding from Korean government (Korea AeroSpace Administration (KASA), grant number RS-2026-25587698).

This research has made use of ``Aladin Sky Atlas'' developed at CDS, Strasbourg Observatory, France \citep{Bonnarel2000}.


\section*{Data Availability}

The data catalogues used in this work are available in the online edition of \citet{Eden2025} in this series: \url{https://doi.org/10.1093/mnras/stae2581}.

The WiggleZ catalogue is available at \url{https://doi.org/10.1093/mnras/stx2963}, and its spectra at \url{https://rdm.uq.edu.au/files/17729f20-50a0-11ec-97fe-7bcb2225c4dd}.

FLASH Survey data are available at \url{https://data.csiro.au/collection/csiro:47965}.



\bibliographystyle{mnras}
\bibliography{HI_absorption_search_bib} 








\bsp	
\label{lastpage}
\end{document}